\documentclass[12pt]{article}
\usepackage{amsmath}
\usepackage{amssymb}
\usepackage{calc}
\usepackage{amsfonts}
\usepackage{graphicx}
\usepackage[ansinew]{inputenc}
\usepackage{color}

\catcode`\@=11
\def\marginnote#1{}
\def\draftlabel#1{{\@bsphack\if@filesw {\let\thepage\relax
   \xdef\@gtempa{\write\@auxout{\string
      \newlabel{#1}{{\@currentlabel}{\thepage}}}}}\@gtempa
   \if@nobreak \ifvmode\nobreak\fi\fi\fi\@esphack}
        \gdef\@eqnlabel{#1}}
\def\@eqnlabel{}
\def\@vacuum{}
\def\draftmarginnote#1{\marginpar{\raggedright\scriptsize\tt#1}}
\def\draft{\oddsidemargin -.5truein
        \def\@oddfoot{\sl preliminary draft \hfil
        \rm\thepage\hfil\sl\today}
        \let\@evenfoot\@oddfoot \overfullrule 3pt
        \let\label=\draftlabel
        \let\marginnote=\draftmarginnote
   \def\@eqnnum{(\theequation)\rlap{\kern\marginparsep\tt\@eqnlabel}
\global\let\@eqnlabel\@vacuum}  }

\def\preprint{\twocolumn\sloppy\flushbottom\parindent 1em
        \leftmargini 2em\leftmarginv .5em\leftmarginvi .5em
        \oddsidemargin -.5in    \evensidemargin -.5in
        \columnsep 15mm \footheight 0pt
        \textwidth 250mmin      \topmargin  -.4in
        \headheight 12pt \topskip .4in
        \textheight 175mm
        \footskip 0pt
        \def\@oddhead{\thepage\hfil\addtocounter{page}{1}\thepage}
        \let\@evenhead\@oddhead \def\@oddfoot{} \def\@evenfoot{} }
\def\signed{
        \def\@oddfoot{\sl \small Adel Bilal -  notes\hfil
        \rm\thepage\hfil\sl \small  \today }
        \let\@evenfoot\@oddfoot 
  \global  }
\def\titlepage{\@restonecolfalse\if@twocolumn\@restonecoltrue\onecolumn
     \else \newpage \fi \thispagestyle{empty}\c@page\z@
        \def\thefootnote{\fnsymbol{footnote}} }

\def\endtitlepage{\if@restonecol\twocolumn \else  \fi
        \def\thefootnote{\arabic{footnote}} \setcounter{footnote}{0}}
\catcode`@=12 \relax
\def\dsp{\displaystyle}
\def\bea{\begin{array}}
\def\eea{\end{array}}
\def\beb{\be\begin{array}{|c|}\hline\\ \dsp}
\def\bebns{\be\begin{array}{|c|}\hline \dsp}
\def\eeb{\\ \\ \hline\end{array}\ee}
\def\eebns{\\ \hline\end{array}\ee}
\def\bab{\be\begin{array}{|ccc|}\hline&&\\ \dsp}
\def\eab{\\ &&\\ \hline\end{array}\ee}
\def\eabns{\\ \hline\end{array}\ee}

\def\bra#1{\left\langle #1\right|}

\def\ket#1{\left| #1\right\rangle}

\def\Im{\mathop{\rm Im}}

\relax
\def\be{\begin{equation}}
\def\ee{\end{equation}}
\def\ba{\begin{eqnarray}}
\def\ea{\end{eqnarray}}
\def\del{\partial}
\def\d{{\rm d}}
\def\tr{\,{\rm tr}\,}
\def\trD{\,{\rm tr}_D\,}

\def\Det{\,{\rm Det}\,}

\def\r{\rho}
\def\a{\alpha}

\def\g{\gamma}

\def\G{\Gamma}
\def\dd{\delta}
\def\D{\Delta}

\def\f{\phi}

\def\p{\psi}

\def\P{\Psi}

\def\m{\mu}
\def\n{\nu}
\def\o{\omega}

\def\l{\lambda}
\def\L{\Lambda}
\def\s{\sigma}

\def\cM{{\cal M}}

\def\cD{{\cal D}}
\def\cR{{\cal R}}

\def\St{{\overline S}}

\def\t{\theta}

\def\dsl{{\del\hskip-2.2mm /}}
\def\dslsmall{{\del\hskip-1.8mm /}}

\def\Nsl{{\nabla\hskip-3.0mm /}}

\numberwithin{equation}{section}

\relax
\renewcommand{\theequation}{\thesection.\arabic{equation}}
\begin{document}
%\signed
\topmargin-2.4cm
%\draft
%\preprint
%
%
%
%
\begin{titlepage}
\begin{flushright}
MIT-CTP/5324 \\
September 2021\\
%\today
\end{flushright}
\vskip 0.5cm

\begin{center}{\Large\bf Effective gravitational action for 2D  massive fermions }\\

\vskip7.mm

{Adel B{\scshape ilal}$^{1}$,
Corinne de L{\scshape acroix}$^{1}$
and Harold E{\scshape rbin}$^{2,3,4}$}
\\

\medskip
$^{1}$\it {Laboratoire de Physique de l'\'Ecole Normale Sup\'erieure\\
PSL University, CNRS, Sorbonne Universit\'e, Universit\'e de Paris\\
24 rue Lhomond, F-75231 Paris Cedex 05, France}

\medskip

$^{2}$\it {Center for Theoretical Physics, MIT, Cambridge, MA 02139, USA}

\medskip

$^{3}$\it{NSF AI Institute for Artificial Intelligence and Fundamental Interactions}

\medskip

$^{4}$\it{Universit\'e Paris Saclay, CEA, LIST, 91191 Gif-sur-Yvette, France}

\end{center}
\vskip .3cm

\begin{center}
{\bf \large Abstract}
\end{center}
\begin{quote}
We work out the effective gravitational action for 2D massive Euclidean  fermions in a small mass expansion. Besides the leading Liouville action, the order $m^2$ gravitational action contains a piece characteristic of the Mabuchi action, much as for 2D massive scalars, but also several non-local terms involving the  Green's functions and Green's functions at coinciding points on the manifold.
\end{quote}

\end{titlepage}
%
%\newpage
%
%
{\parskip -0.3mm
\small{\tableofcontents}}

\setcounter{footnote}{0} 
\setlength{\baselineskip}{.55cm}

\newpage
\setcounter{page}{1}

%

%%%%%%%%%%%%%%%%%%%%%%%%%%%%%%%%%%%%%
\setcounter{section}{0}

%%%%%%%%%%%%%%%%%%%%%%%%%%%%%%%%%%%%%
\section{Introduction\label{intro}}
%%%%%%%%%%%%%%%%%%%%%%%%%%%%%%%%%%%%%
The effective gravitational action for a given matter system coupled to gravity is defined in terms of the matter partition functions $Z_{\rm mat}[g]$ computed in a fixed metric $g$ on a given manifold $\cM$ as
\be\label{Sgravgen}
\exp\big(-S_{\rm grav}[g,\hat g] \big) = \frac{Z_{\rm mat}[g]}{Z_{\rm mat}[\hat g]}\ .
\ee
By this definition, the effective gravitational action necessarily depends on two different metrics. We may consider $\hat g$ as a reference metric, and in particular in two dimensions we may assume that $g$ and $\hat g$ are related by a conformal factor as $g=e^{2\s}\hat g$.
The definition \eqref{Sgravgen} implies that $S_{\rm grav}$ satisfies a cocycle identity
\be\label{cocycle}
S_{\rm grav}[g_1,g_2] + S_{\rm grav}[g_2,g_3]=S_{\rm grav}[g_1,g_3] \ .
\ee
The  best-known example is the effective gravitational action for conformal matter coupled to 2D gravity which is the Liouville action  \cite{Liouville1}
\be\label{Liouvaction}
S_L[g,\hat g] \equiv S_L[\hat g,\s]=\int\d^2 x \sqrt{\hat g} \, \big( \s \D_0\s + R_0 \s\big) 
, \quad g=e^{2\s} \hat g\ ,
\ee
and
\be\label{Liouvillegrav}
S_{\rm grav}[g,\hat g]=-\frac{c}{24\pi} S_L[g,\hat g] \ ,
\ee
$c$ being the central charge of the conformal matter system.
Another, even simpler example of a gravitational action satisfying this cocycle identity is the ``cosmological constant action"
\be\label{cosmolconst}
S_c[g,\hat g]=\m_0  \int \d^2 x (\sqrt{g} -\sqrt{\hat g})=\m_0 (A-A_0) \ .
\ee
In general, this action must also be present as a counterterm to renormalize the divergences that are present in \eqref{Sgravgen}, in addition to $S_L$.

Two-dimensional gravitational actions other than the Liouville or cosmological constant actions can be constructed and have been studied  in the mathematical literature, like the Mabuchi and Aubin-Yau actions \cite{Mabuchi,AubinYau}. These latter functionals   involve not only the conformal factor $\s$ but also  the K\"ahler potential $\f$ and do admit generalizations to higher-dimensional  K\"ahler manifolds. In the mathematical literature they  appear in relation with the characterization of constant scalar curvature metrics \cite{AubinYau}.
Their r\^oles as two-dimensional gravitational actions in the sense of \eqref{Sgravgen} have been highlighted in \cite{FKZ}. In particuler, ref.~\cite{FKZ} has initiated the study of the metric dependence of the partition function of non-conformal matter like a massive scalar field and shown that  the gravitational action defined by \eqref{Sgravgen} contains these Mabuchi and Aubin-Yau actions  as first-order corrections (first order in $m^2 A$ where $m$ is the mass and $A$ the area of the Riemann surface) to the Liouville action.
The partition function of quantum gravity at fixed area, with a gravitational action being a combination of the Liouville and Mabuchi actions, has been studied at one loop in \cite{BFK} and at  two and three loops in \cite{BL}. The effective gravitational action\footnote{
Note that at present we call $S_{\rm grav}[g,\hat g]$ what was called $S_{\rm grav}[\hat g,g]$ in refs \cite{FKZ,BL,BL2,BdL2}.
} 
for such massive scalar fields including higher-order corrections in $m^2$ has been studied in \cite{BL2}, and on manifolds with boundaries in \cite{BdL2}. A rigourous mathematical construction of the functional integral based on the coupling of the Liouville and Mabuchi actions has since been obtained in ref \cite{LRV} by means of probabilistic tools.  Further properties of the Mabuchi action were studied in \cite{dL1,dLES}.

References \cite{FKZ,BL2,BdL2} all studied the effective gravitational action of a two-dimensional massive scalar field, and a natural question  to ask was what happens for two-dimensional massive fermions. As we will briefly discuss below, the most interesting case corresponds to a massive ``Majorana" fermion.\footnote{
This is a 2D fermion with a Majorana-type mass term. Classically, at the level of the Dirac equation, one can impose a Majorana condition, but the eigenfunctions of the corresponding (purely imaginary) Dirac operator $D$ necessarily are complex. One might argue whether it is then appropriate to talk about Majorana fermions in the quantum theory. However, we can obtain the partition function in terms of the eigenvalues of the (real) squared operator $D^2$ whose eigenfunctions can all be chosen to be real.}
While this seemed to be a simple generalisation of the case of the scalar field, it actually turned out to be technically quite more involved. It is the purpose of the present paper to study this effective gravitational action for massive fermions. Much as for the scalar case, the renormalised Green's functions at coinciding points play an important role. At present it will be useful to study both, the Green's function of the massive Dirac operator $D$ and the Green's function of the squared  operator $D^2$. The study of these operators and the associated (local) zeta-functions and (local) heat kernels will occupy a major part of these notes. The order $m^2$ terms of the effective gravitational action can then be expressed as an integral over the manifold of these Green's functions at coinciding points. A detailed knowledge of their zero-modes (for zero mass) is required to perform the small-mass expansion correctly. In this paper we will then restrict to spherical topology where the Dirac operator has no zero-mode (even for zero mass) and we can reliably obtain the small mass expansion up to and including the order $m^2$ terms. Besides the leading Liouville term, at order $m^2$ we get a cosmological constant action, and a local $\int\sqrt{\hat g}\, \s e^{2\s}$ term characteristic of the Mabuchi action, as well as some further non-local terms involving the Green's functions. Despite some effort, we could not express the latter in terms of purely local quantities like the conformal factor $\s$ and the K\"ahler potential. Similar non-local terms also appeared in the scalar case at order $m^4$ and it seems that in the present fermionic case they already are unavoidable at order $m^2$.
We  plan to return to the case of general topology with a detailed account of the role of the zero-modes in a future publication.

This paper is organised as follows. In the next section we introduce the relevant differential operators : massive Dirac operator $D$, its square $D^2$, and the scalar and spinoral Laplacians. We discuss their eigenvalue problems (with a specific look at the example of the torus) and how the eigenvalues transform under conformal changes of the metric. In section 3 we define the gravitational action precisely in terms of these eigenvalues and how it is related to the corresponding zeta-functions. In section 4 we embark on a detailed study of the different Green's functions, local zeta-functions and local heat kernels, and how they vary under conformal changes of the metric.  So far, all our discussion is valid for all 2D topologies. All this is put together in section 5 to obtain the effective gravitational action. The first two subsection are still for general topology, but then we specify to spherical topology to be able to do the small $m$-expansion and identify the order $m^2$ contributions to the effective action, and in particular display the Mabuchi-type action.

\newpage
%%%%%%%%%%%%%%%%%%%%%%%%%%%%%%%%%%%%%
\section{The Dirac operator\label{Dirac}}
%%%%%%%%%%%%%%%%%%%%%%%%%%%%%%%%%%%%%
\setcounter{equation}{0}

%%%%%%%%%%%%%%%%%%%%%%%%%%%%%%%%%%%%%
\subsection{Flat space : $\g$-matrices and Dirac operator}

In D=2 Euclidean dimensions, $\m=1,2$, we choose two hermitian $\g$-matrices.  We can choose them to be both real, i.e. $\s_x$ and $\s_z$. Thus
\be\label{gammachoice}
\g^1=\s_x \ ,  \quad \g^2=\s_z 
%\quad \Rightarrow\quad (\g^\m)^\dag=\g^\m \quad , \quad (\g^\m)^T=\g^\m \quad , \quad (\g^\m)^*=\g^\m 
\ .
\ee
The chirality matrix then is 
\be\label{gamma5}
\g_*=i\g^1\g^2=\s_y \quad \Rightarrow\quad \g_*^\dag=\g_* \quad , \quad \g_*^T=\g_*^*=-\g_* \ ,
\ee
and, of course, $\g_*$ anticommutes with $\g^\m$. The generator of ``Lorentz" transformations, i.e. of SO(2) is $\frac{i}{4}[\g^1,\g^2]=\g_*$ and the  representation of a finite rotation by an angle $\a$ on a spinor $\p$ is given by the real matrix
$D(\a)=e^{i\a\g_*/2}$.
One easily shows that $\p^*$, $\g_*\p$ and $\dsl\p$ all transform\footnote{
The precise statement for $\dsl\p$  is, of course : if $\p'(x')=D(\a)\p(x)$ then $\g^\m \frac{\del}{\del {x'}^\m}\p'(x')=D(\a) \g^\n \frac{\del}{\del x^\n} \p(x)$.
} 
exactly as $\p$. Hence it makes sense to impose the Majorana condition $\p^*=\p$.

For anticommuting Majorana spinors a mass term like  $\int \p^\dagger m \p =\int m \p^{\tt T}\p$ vanishes. We can, however, introduce a non-vanishing, real mass-term as $\int \p^\dag m\g_*\p$, so that the action reads
\be\label{Majaction}
S=\int \p^\dag ( i\dsl + m\g_*)\p \ .
\ee
The corresponding Dirac operator
\be\label{flatDirop}
D=i\dsl + m\g_*
\ee 
is hermitian and squares to $D^2=-\del_\m\del^\m + m^2$ which incorporates the correct Euclidean continuation of the mass-shell condition $p_\m p^\m + m^2=0$. For  (complex) Dirac spinors  one might also contemplate an ordinary mass term with action 
$S=\int \p^\dag ( i\dsl + m)\p$. However, the square of the corresponding Dirac operator $(i\dsl + m)^2 = -\del_\m\del^\m + m^2+2im \dsl$ does {\it not} correspond to anything simple or physical. (It is $(i\dsl-m) (i\dsl +m)$ that instead gives the mass-shell condition.) For this reason, we will focus on the action \eqref{Majaction} and corresponding Dirac operator \eqref{flatDirop}.

Note that the Dirac operator \eqref{flatDirop}
is a purely imaginary hermitian differential operator and hence $(i\dsl + m\g_*)\p=0$ admits real solutions $\p$. However, the corresponding eigenvalue problem 
\be\label{evproblemflat}
(i\dsl + m\g_*)\p_n=\l_n\p_n
\ee
clearly cannot admit real solutions $\p_n$ (since $\l_n$ are real) and on must take the $\p_n$ to be complex. Taking the complex conjugate of \eqref{evproblemflat}, we see that $\p_n^*$ also is an eigenfunction but with eigenvalue $-\l_n$.
If we let
\be\label{psichiphiflat}
\p_n=\frac{1}{\sqrt{2}}(\chi_n+i \f_n)\ ,
\ee
then taking the real and imaginary part of \eqref{evproblemflat} one gets
\be\label{evphichiflat}
(i\dsl + m\g_*)\chi_n=i \l_n\f_n
\quad , \quad
(i\dsl + m\g_*)\f_n=-i \l_n\chi_n \ ,
\ee
and
\be\label{ev2phichiflat}
(i\dsl + m\g_*)^2\chi_n=\l_n^2\chi_n
\quad , \quad
(i\dsl + m\g_*)^2\f_n= \l_n^2\f_n \ .
\ee

%%%%%%%%%%%%%%%%%%%%%%%%%%%%%%%%%%%%%
\subsection{Curved space : $\g$-matrices, Dirac operator and spinorial Laplacian}

The above Dirac-matrices (to be denoted $\g^a$) and Dirac operator are those in flat space. In curved space we have
$\g^\m=E^\m_a \g^a$ and the spinorial covariant derivative is 
\be\label{spinorcovder}\nabla_\m=\del_\m - \frac{i}{2}\o_\m^{ab}\frac{i}{4}[\g^a,\g^b]
=\del_\m +\frac{1}{4}\o_\m^{ab}\g^{ab}=\del_\m -\frac{i}{2}\o^{12}_\m \g_* \ .
\ee
We let
\be\label{omegamudef}
\o_\m\equiv 2\o_\m^{12} \quad  \Rightarrow\quad
\nabla_\m=\del_\m -\frac{i}{4}\o_\m\g_* \ .
\ee
More generally when acting on a spinorial object that also carries vector indices, one has $\nabla_\m=\del_\m -\frac{i}{4}\o_\m\g_*+\G_\m$. To avoid confusion we call $\nabla_\m^{\rm sp}$ the spinor covariant  derivative defined in \eqref{spinorcovder}, \eqref{omegamudef}~:
\be\label{spinorcovderbis}
\nabla_\m^{\rm sp}=\del_\m -\frac{i}{4}\o_\m\g_* \ .
\ee
Note that since $\g_*=\s_y$ is purely imaginary, $\nabla_\m^{\rm sp}$ is a real (anti-hermitian) differential operator.
Next we define
\be\label{nablaslash}
\Nsl = \g^\m\, \nabla_\m^{\rm sp}=\g^a E_a^\m\, \nabla_\m^{\rm sp}\ .
\ee
It is important to note that $\Nsl\p$ transforms as $\p$ and hence $\Nsl\, \Nsl\p= \g^\m  \nabla_\m^{\rm sp} \g^\n  \nabla_\n^{\rm sp} \p$. However, this is {\it not} $\g^\m  \g^\n \nabla_\m^{\rm sp}  \nabla_\n^{\rm sp}\p$. Instead one has 
$\nabla_\m \g^a=\o_\m^{ab}\g^b$ and $\nabla_\m\g^\n=E^\n_a \nabla_\m \g^a=E_a^\n \o_\m^{ab}\g^b= -\g^a\o_\m^{ab}E_b^\n$. 
But using $0=\nabla_\m E_a^\n=\del_\m E_a^\n +\G^\n_{\m\r}E^\r_a +\o_{ab}E_b^\n$ we get consistently
\be\label{gammacovder}
\nabla_\m\g^\n=\del_\m\g^\n +\G^\n_{\m\r}\g^\r\ .
\ee
One can then consistently show that $\nabla_\m \Nsl\p = \nabla_\m^{\rm sp} \Nsl\p$ and
\be\label{Nslsquare}
\Nsl\, \Nsl \p \equiv (\g^\m \nabla_\m) (\g^\n \nabla_\n) \p = \g^\m \g^\n \nabla_\m \nabla_\n\p
=  \g^\m \g^\n \Big( \nabla_\m^{\rm sp} \nabla_\n^{\rm sp}  -\G_{\m\n}^\r \nabla_\r^{\rm sp}\Big)\p\ .
\ee
This is as if $\nabla_\m \g^\n$  vanishes, but as we have seen above this is not the case. What one really has is
\be\label{gammanablacomm}
[\nabla_\m,\g^\n]=0 \quad , {\rm i.e.}\quad \nabla_\m ( \g^\n \ldots) = \g^\n \nabla_\m \ldots \ .
\ee
Next, using $\g^\m\g^\n=g^{\m\n}+\g^{\m\n}$ and $\nabla_\m \nabla_\n\p-\nabla_\n \nabla_\m\p = \frac{1}{4} R_{\m\n}^{ab}\g^{ab}\p$ we get
\be\label{Nsl2}
\Nsl\, \Nsl \p =g^{\m\n}\nabla_\m\nabla_\n \p +\frac{1}{8} \g^{cd}R^{ab}_{cd}\g^{ab}\p  \ .
 \ee
In two dimensions $\g^{cd}R^{ab}_{cd}\g^{ab}$ simplifies to $-2\cR$ where $\cR$ is the scalar Ricci curvature. Thus
\be\label{Nsl3}
\Nsl\, \Nsl \p =g^{\m\n}\nabla_\m\nabla_\n \p -\frac{1}{4} \cR \p = (\D_{\rm sp} -\frac{1}{4}\cR)\p \ ,
 \ee
 where we have introduced the spinorial Lapacian :
 \be\label{spinorLapl}
 \D_{\rm sp} \p= g^{\m\n}\nabla_\m\nabla_\n \p = g^{\m\n} \Big( \nabla_\m^{\rm sp} \nabla_\n^{\rm sp}  -\G_{\m\n}^\r \nabla_\r^{\rm sp}\Big)\p \ .
 \ee
Note that $\nabla_\m^{\rm sp}$ is a $2\times 2$-matrix differential operator, and so is this spinorial Laplacian. More precisely, we have a piece proportional to the identity matrix and involving the scalar Laplacian, and a piece proporional to $\g_*$:
\be\label{spinorLapl2}
 \D_{\rm sp} \p= \Big( \D_{\rm scalar} -\frac{1}{16}\o^\m\o_\m \Big)\p 
 -\frac{i}{4}\, \g_*\, \Big( (\nabla_\m \o^\m) + 2 \o^\m \del_\m \Big)\p \ .
 \ee
Finally, we define the purely imaginary, hermitian Dirac operator as the curved-space generalisation of \eqref{flatDirop}~:
\be\label{Dop}
D=i\Nsl+m\g_* \ .
\ee
Then
\be\label{Dsquared}
D^2=-\Nsl\, \Nsl + m^2 = -\D_{\rm sp} +\frac{1}{4} \cR + m^2 \ .
\ee

Again, as in flat space, the eigenvalue problem of $D$
\be\label{evproblem}
D \p_n\equiv (i\Nsl + m\g_*)\p_n=\l_n\p_n \ ,  
\ee
($n\ge 0$) cannot admit real solutions, and $\p_n^*$ is eigenfunction with eigenvalue $-\l_n$~:
\be\label{evproblem2}
D \p_n^*=-\l_n \p_n^* \ .
\ee
For all $\l_n\ne 0$, within any couple $(\p_n,\p_n^*)$ we decide to call $\p_n$ the eigenfunction with the positive eigenvalue. With this convention we always have 
\be\label{lambdannonnegative}
\l_n\ge 0\ .
\ee
Again we let
\be\label{psichiphi}
\p_n=\frac{1}{\sqrt{2}}(\chi_n+i \f_n)\ ,
\ee
and get
\be\label{evphichi}
D \chi_n=i \l_n\f_n
\quad , \quad
D \f_n=-i \l_n\chi_n \ ,
\ee
while, of course
\be\label{ev2phichi}
D^2\chi_n=\L_n\chi_n
\quad , \quad
D^2\f_n= \L_n\f_n 
\quad , \quad
D^2\p_n=\L_n\p_n 
\quad , \quad \L_n= \l_n^2\ .
\ee

The natural inner product is
\be\label{innerprod}
(\P_1,\P_2)=\int\d^2 x \sqrt{g}\, \P_1^\dag \P_2 \ .
\ee
With respect to this inner product $i\Nsl$ and $D$ are hermitian operators and hence the $\p_n$ corresponding to different $\l_n$ are orthogonal. Similarly, for the $\p_n^*$. Also all $\p_n^*$ are orthogonal to all $\p_k$ (as long as $\l_n\ne 0$). This, together with the normalisation conditions is equivalent to
\be\label{orthonomality}
(\chi_n,\chi_k)=\dd_{nk} \quad , \quad
(\f_n,\f_k)=\dd_{nk} \quad , \quad
(\chi_n,\f_k)=0 \quad (\l_n\ne 0)\ .
\ee

\noindent
{\bf Remarks : }
Let us make a few obvious remarks.
Note that the eigenfunctions of $D$ (the $\p_n$ and $\p_n^*$) are automatically eigenfunctions of $D^2$, but the converse is not necessarily true as is examplified by the $\chi_n$ and $\f_n$. What is true is that within any eigenspace of $D^2$ with eigenvalue $\L_n=\l_n^2$ one can find linear combinations (corresponding precisely to the $\p_n$ and $\p_n^*$) that are eigenfunctions of $D$ with eigenvalues $+\l_n$ and $-\l_n$.

Since $\nabla_\m^{\rm sp}$ commutes with $\g_*$, it is clear that $\Nsl$ anticommutes with $\g_*$. Then, for $m\ne 0$, $D$ has no simple (anti)commutation relation with $\g_*$ and we cannot have eigenfunctions of $D$ of definite chirality (i.e. being also eigenfunctions of $\g_*$). For $m=0$, $D$ anticommutes with $\g_*$, so that $\g_*\p_n$ is eigenfunction of $D$ with eigenvalue $-\l_n$. Thus for $\l_n\ne 0$, $\p_n$ and $\g_*\p_n$ necessarily are orthogonal. For $\l_n=0$, however, one can always choose a basis of definite chirality eigenfunctions.\footnote{If $\p_{0,i}$ already has definite chirality, nothing is to be done. If not, then $(1\pm\g_*)\p_{0,i}/2$ are both non-vanishing and have definite chirality $\pm1$.}
 As is well known, the difference of positive and negative chirality zero-modes of $D$ is called its index.

For $D^2$ the situation is simpler : it is clear from  \eqref{spinorLapl2} and \eqref{Dsquared} that $D^2$ commutes with $\g_*$, and one can then take the eigenfunctions of $D^2$ to have definite chirality. From the discussion of the previous paragraph it is then clear that in general these definite chirality eigenfunctions of $D^2$ are {\it not} eigenfunctions of $D$.

Finally, one has
\be\label{positivity}
\l_n^2-m^2=(\p_n,(D^2-m^2)\p_n)=(\p_n,(i\Nsl)^2\p_n)=(i\Nsl\p_n,i\Nsl\p_n) \ge 0 \ .
\ee
We see that $\l_n^2 > m^2$, unless $i\Nsl \p_n=0$ in which case $\l_n^2=m^2$. In any case, for $m\ne 0$ we have $\l_n^2>0$ and there are no zero-modes of $D$.
Similarly, we have 
\be\label{Spinorlaplcond}
(\p_n,(-\D_{\rm sp})\p_n)=-\int\d^2 x\sqrt{g}\, \p_n^* g^{\m\n}\nabla_\m \nabla_\n\p_n
=\int\d^2 x\sqrt{g} (\nabla_\m^{\rm sp}\p_n)^* g^{\m\n} \nabla_\n^{\rm sp}\p_n \equiv a^2 \ge 0 \ .
\ee
It follows  from \eqref{Dsquared} that if we choose a constant curvature metric  $\cR=\bar\cR$, we have
\be\label{meancurvrel}
\l_n^2-m^2 = a^2 +\frac{1}{4} \bar\cR \ ,
\ee
so that on the sphere with $\cR>0$ we have $\l_n^2 > m^2$ for all $n$, i.e. there are no zero-modes even for $m=0$.

%%%%%%%%%%%%%%%%%%%%%%%%%%%%%%%%%%%%%
\subsection{The example of the flat torus}\label{flattorus}
 
It is useful to have at least one specific example of fermionic eigenfunctions. The simplest case is the flat torus with each of its periods being $2\pi$. Of course, being flat, we will not be able to appreciate any effects of curvature.
As is well known, for fermions we may consider different spin structures, i.e. periodic or anti-periodic boundary conditions around one or the other circle of the torus, leading to four different spin structures. Here we will only investigate the doubly periodic boundary conditions, the other spin structures can be treated in complete analogy. Note that it is only for the doubly periodic boundary conditions that the zero-modes are present.

The relevant Dirac operator then is the one given in \eqref{evproblem}, i.e.
\be\label{flatD}
D=i\dsl+m\g_* = i\s_x\del_1 + i \s_z \del_2 + \s_y m =i\, \begin{pmatrix} \del_2 & \del_1 - m\\ \del_1 + m & - \del_2 \end{pmatrix} \ .
\ee
We have $D^2=-(\del_1^2+\del_2^2) + m^2$ so that $D^2 \p_n=\l_n^2\p_n$ yields
\be\label{lambdatorus}
\l_n^2\equiv \l_{\vec{n}}^2=n_1^2 + n_2^2 + m^2 \quad , \quad \p_n\sim e^{i n_1 x^1 + i n_2 x^2}\ .
\ee
The complex eigenfunctions of the Dirac operator $D$ then are
\be\label{eigentorus}
\p_{\vec{n}}(x^1,x^2)=\begin{pmatrix} a\\ b \end{pmatrix} e^{i n_1 x^1 + i n_2 x^2} \ ,
\ee
where $a$ and $b$ of course also depend  on the integers $n_1$ and $n_2$ but not on $x^1$ and $x^2$. We denote by $\l_{\vec{n}}$ the positive square root of $\l_{\vec{n}}^2$\ :\ 
$\l_{\vec{n}}=+\sqrt{n_1^2 + n_2^2 + m^2}$. Then
$D\p_n=\l_n\p_n$ gives
\be\label{systeq}
-(n_1+im) b=(\l_{\vec{n}}+n_2)a \quad , \quad -(n_1-im) a=(\l_{\vec{n}}-n_2)b \ .
\ee
This is easily solved, up to a common normalisation, which we choose such that $\int \d^2x\,  \p_{\vec{n}}^\dag \p_{\vec{n}}=1$~:  
\be\label{eigentorussol}
\p_{\vec{n}}(\vec{x})
=\frac{1}{2\pi \sqrt{2\l_{\vec{n}}(\l_{\vec{n}}+n_2)}} 
\begin{pmatrix} n_1+im \\ -\l_{\vec{n}}-n_2 \end{pmatrix} 
e^{i n_1 x^1 + i n_2 x^2} \ .
\ee
Of course, since $\l_{\vec{n}}$ only depends on $n_1^2+n_2^2$, $\p_{\vec{n}}$ and $\p_{-\vec{n}}$ have the same eigenvalue $\l_{\vec{n}}$. One also easily checks that
\be\label{psinstar}
\p_{\vec{n}}^*= \p_{-\vec{n}}\big\vert_{\l_{\vec{n}}\to -\l_{\vec{n}}} \ ,
\ee
which confirms that $\p_{\vec{n}}^*$ is indeed an eigenfunction of $D$ with eigenvalue $ -\l_{\vec{n}}$. Thus the complete set of eigenfunctions is the set of all $\p_{\vec{n}}$ and all $\p_{\vec{n}}^*$. Note the degeneracy of the eigenvalues. For non-zero $n_1$ and $n_2$, $\pm n_1$ and $\pm n_2$ all yield the same $\l_{\vec{n}}$, and thus the eigenvalues of $D$ are (at least) four-fold degenerate. (In the present example where both circles have the same radius, there is a further degeneracy under the exchange of $n_1$ and $n_2$.) Since for $D^2$, $\p_n$ and $\p_n^*$ correspond to the same eigenvalue $\L_n=\l_n^2$, generically the eigenvalues of $D^2$ are (at least) eight-fold degenerate.

Note that for $m\ne 0$ there are no zero-modes, while for $m=0$ the zero-modes correspond to $n_1=n_2=0$. However, the form of 
 \eqref{eigentorussol} is not directly useful as it is indeterminate in this case. We may instead first take to $n_1=n_2=0$ and then consider the limit $m\to 0$ which gives the two zero-modes
\be\label{toriszeromodes}
\p_{\vec{0}}=\frac{-1}{2\pi\sqrt{2}} \begin{pmatrix} -i \\ 1 \end{pmatrix} \quad , \quad
\p_{\vec{0}}^*=\frac{-1}{2\pi\sqrt{2}} \begin{pmatrix} i \\ 1 \end{pmatrix}  \ .
\ee
Note that they happen to be eigenmodes of $\g_*$ :
\be\label{zeromgammmastar}
\g_* \p_{\vec{0}}= \p_{\vec{0}} \quad , \quad \g_* \p_{\vec{0}}^*= -  \p_{\vec{0}} \ .
\ee
 
For the other spin structures, the $n$ corresponding  to a circle with anti-periodic boundary conditions are half-integer and, obviously, then there are no zero-modes even for $m=0$.

%%%%%%%%%%%%%%%%%%%%%%%%%%%%%%%%%%%%%%%%%%%%%%%%%%%%%
\subsection{Conformal changes}\label{confsec}

We will consider conformal changes between a ``reference" metric / vielbein and a conformally rescaled metric / vielbein, related as
\be\label{confresc}
g_{\m\n}=e^{2\s} \hat g_{\m\n} \quad , \quad e^a_\m=e^\s \hat e^a_\m \quad , \quad E_a^\m=e^{-\s} \hat E_a^\m \ ,
\ee
where the conformal factor $\s$ depends on the space-coordinates. It follows from $\d e^a+\o^{ab}e^b=0$ ($e^a=e^a_\m \d x^\m$, $\o^{ab}=\o^{ab}_\m \d x^\m$) that
\be\label{omegaresc}
\o^{ab}_\m=\hat\o^{ab}_\m +(\hat e^a_\m \hat E^{b\l} - \hat e^b_\m \hat E^{a\l})\del_\l\s \ .
\ee
This yields
\be\label{spinorderconf}
\nabla_\m^{\rm sp} = \hat \nabla_\m^{\rm sp} -\frac{i}{2} \big(\hat e^1_\m \hat E^{2\l}-\hat e^2_\m \hat E^{1\l} \big) \del_\l\s \ .
\ee
Using $\g^\m=E^\m_a\g^a=e^{-\s}\hat\g^\m$, it is then a straightforward exercice to show
\be\label{Nslconf}
\Nsl = e^{-\s} \big( \hat\Nsl +\frac{1}{2} \hat\g^\l\del_\l \s\big) =e^{-3\s/2}\,\hat\Nsl \ e^{\s/2} \ .
\ee
For an infinitesimal variation $\dd\s$  of the conformal factor this yields
\be\label{Nslvar}
\dd \Nsl= -\dd\s \Nsl + \frac{1}{2} \g^\l\del_\l \dd\s 
\quad \Rightarrow \quad
\dd D= -\dd\s (D-m\g_*) +  \frac{i}{2} \g^\l\del_\l \dd\s\ .
\ee
Next, for the Christoffel symbols one has
\be\label{Christvar}
g^{\m\n}\G_{\m\n}^\r=e^{-2\s} \hat g_{\m\n} \hat \G_{\m\n}^\r \ .
\ee
To work out the variation of $D^2$ one can then either use \eqref{Nslconf} or \eqref{spinorderconf} together with \eqref{spinorLapl}
 and \eqref{Christvar}. In any case,
\be\label{D2confscal}
D^2-m^2 = e^{-2\s} \Big( D^2-m^2   + \hat\g^{\m\n} \del_\m\s \hat\nabla_\n^{\rm sp} -\frac{1}{2} (\hat\D_{\rm scalar} \s) +\frac{1}{2} \hat g^{\m\n} (\del_\m\s)(\del_\n\s )\Big) \ ,
\ee
or for an infinitesimal variation of $\s$ :
\be\label{D2confscal2}
\dd D^2= -2\dd \s ( D^2-m^2)  +(\del_\m \dd\s )\g^{\m\n} \nabla_\n^{\rm sp}   - \frac{1}{2} (\hat\D_{\rm scalar} \dd\s)  \ .
\ee 

\newpage
%%%%%%%%%%%%%%%%%%%%%%%%%%%%%%%%%%%%%%%%%
\section{The gravitational action}

We define the matter partition function for fermionic matter with action $S=\int\p^\dag(i\Nsl+m\g_*)\p$ on a two-dimensional manifold with metric $g$ as
\be\label{Zg}
Z_{\rm mat}[g]= \int \cD \P \exp\left( - \int\d^2 x \sqrt{g}\, \P^\dag D_g \P\right) \ ,
\ee
where we wrote $D_g$ to insist that this is the Dirac operator $D$ for the metric $g$ (and corresponding vielbein $e$ and spin connection $\o$). One expands $\P$ on a complete set of eigenmodes of $D$:
\be\label{eigenexp}
\P(x)=\frac{1}{\sqrt{\m}}\Big( \sum_n (b_n \chi_n(x) + c_n \f_n(x)) +\sum_i d_i \p_{0,i}(x) \Big) \ ,
\ee
where the first sum does not include the zero modes $\p_{0,i}$. The eigenfunctions $\chi_n,\ \f_n$ and $\p_{0,i}$ are real, commuting functions, orthonormalized as discussed above. $\m$ is an arbitrary mass scale we introduce so that the
the anticommuting coefficients $b_n, \ c_n, d_i$ are dimensionless.\footnote{
Indeed, from the normalisation condition of the eigenmodes one sees that the $\chi_n$ and $\f_n$ have engineering dimension one, i.e. $\chi_n\sim\f_n\sim \m$, and
since $\P$ must have dimension $\frac{1}{2}$ so that the action 
$ \int\d^2 x \sqrt{g}\, \P^\dag D_g \P$ is dimensionless, we see from \eqref{eigenexp} that  $b_n, \ c_n, d_i$ are indeed dimensionless.
} 
It follows that
\be\label{actionexp}
 \int\d^2 x \sqrt{g}\, \P^\dag D_g \P= 2 i \sum_n \frac{\l_n}{\m} c_n b_n \ ,
 \ee
 where of course only the non-zero modes contribute. The functional integral measure is defined in terms of grassmann integrals over these coefficients $b_n$ and $c_n$, so that
\be\label{Zg2}
Z_{\rm mat}[g]= \int \prod_n \d b_n \d c_n \exp\left( -2i \sum_n \frac{\l_n}{\m} c_n b_n  \right) \ = {\cal N} \prod_n \frac{\l_n}{\m}\ .
\ee
(again, the product is only over all strictly positive eigenvalues $\l_n$.

The gravitational action was defined by \eqref{Sgravgen}, i.e.
$
S_{\rm grav}[g,\hat g]=-\ln \frac{Z_{\rm mat}[g]}{Z_{\rm mat}[\hat g]} \ ,
$
so that
\be\label{gravac2}
S_{\rm grav}[g,\hat g]=-\ln \prod_n \frac{\l_n[g]}{\m}+ \ln  \prod_n \frac{\l_n[\hat g]}{\m} \ .
\ee
We may also rewrite this in terms of the determinant of $D^2$.
The operator $D^2$ has eigenvalues $\l_n^2$ for the $\chi_n$ and $\l_n^2$ for the $\f_n$, so that
\be\label{D2det}
\Det' D^2 =\big( \prod_n \l_n^2 \big)^2\ ,
\ee
and 
\be\label{gravac3}
S_{\rm grav}[g,\hat g]=-\frac{1}{4}\ln \Det' D^2[g] + \frac{1}{4}\ln \Det' D^2 [\hat g] \ .
\ee

All these determinants and products of eigenvalues are, of course, ill-defined and have to be regularised. We will use the standard tool of regularisation via the corresponding zeta-functions. The zeta-function of the operator $D^2$ is
\be\label{ZetaD2}
\zeta(s)=2\sum_n (\l_n^2)^{-s} \ ,
\ee
since every eigenvalue $\l_n^2>0$ occurs once with eigenfunction $\chi_n$ and once with $\f_n$. Also, the sum obviously does not include any zero eigenvalue in case there are zero-modes. Standard manipulations give for the derivative
\be\label{zetader}
\zeta'(0)=-2\sum_n \ln \l_n^2 \quad \Rightarrow \quad
-2\ln \prod_n \frac{\l_n^2}{\m^2} =\zeta'(0)+\ln \m^2 \zeta(0) \ ,
\ee
It follows that the (regularised) gravitational action is
\be\label{gravac4}
S_{\rm grav}[g,\hat g]=\frac{1}{4} \Big( \zeta_g'(0) +\ln\m^2 \zeta_g(0)\Big) 
- \frac{1}{4} \Big( \zeta_{\hat g}'(0) +\ln\m^2 \zeta_{\hat g}(0)\Big) \ .
\ee

We want to determine this gravitational action for $g_{\m\n}(x)=e^{2\s(x)} (\hat g)_{\m\n}(x)$. Our strategy will be to first determine $\dd S_{\rm grav}$ for infinitesimal $\dd \s$ and then ``integrate" this variation to obtain $S_{\rm grav}[g,\hat g]$. Obviously, the variation of the gravitational action is given in terms of the variation of the zeta-function $\zeta(s)$ around $s=0$. To obtain this, we need to study the variations of the eigenvalues $\l_n$ under a corresponding variation of the Dirac operator $D$. This leads us to the study of the Green's functions, local zeta functions and local heat kernels and their variations which is the subject of the next section.

%\newpage
%%%%%%%%%%%%%%%%%%%%%%%%%%%%%%%%%%%%%%%%%%%%%%%%%%%%%%%%%
\section{Green's functions, heat kernels and zeta-function}

%%%%%%%%%%%%%%%%%%%%%%%%%%%%%%%%%%%%%%%%%%%%%%%%%%%%%%%%%
\subsection{Definitions and basic relations}

Throughout this section we assume that $m\ne 0$ so that there are no zero-modes of $D$. (We could also include the case $m=0$ if the manifold is a sphere.) Recall that we had defined $\p_n$ and $\p_n^*$ (and hence $\chi_n$ and $\f_n$) such that $\l_n\ge 0$. Thus throughout this section we can assume $\l_n>0$.

To begin with, note that
\be\label{completeness}
\sum_n \big(\chi_n(x)\chi_n^\dag(y)+\f_n(x)\f_n^\dag(y)\big) = \sum_n \big(\p_n(x)\p_n^\dag(y)+\p^*_n(x)(\p_n^*)^\dag(y)\big)  = \frac{\dd(x-y)}{\sqrt{g}}\ {\bf 1}_{2\times 2}\ .
\ee

%%%%%%%%%%%%%%%%%%%%%%%%%%%%%%%%%%%%%%%%%%%%%%%%%%%%%%%%%
\subsubsection{Local zeta-functions}

Next, we define two local zeta-functions $\zeta_+(s,x,y)$ and $\zeta_-(s,x,y)$ as
\ba\label{Zetadefs}
\zeta_+(s,x,y)&=&\sum_n \l_n^{-2s} \big(\chi_n(x)\chi_n^\dag(y)+\f_n(x)\f_n^\dag(y)\big) \ ,\nonumber \\
\zeta_-(s,x,y)&=&\sum_n \l_n^{-2s} \big(\chi_n(x)\f_n^\dag(y)-\f_n(x)\chi_n^\dag(y)\big) \ ,
\ea
where $x=(x^1,x^2)$ and $y=(y^1,y^2)$ denote points on the manifold.  Note that these local zeta-functions are real $2\times 2$-matrices.
They can also be rewritten in terms of the $\p_n$ and $\p_n^*$ as
\ba\label{Zetadefs2}
\zeta_+(s,x,y)&=&\sum_n \l_n^{-2s} \big(\p_n(x)\p_n^\dag(y)+\p^*_n(x)(\p_n^*)^\dag(y)\big) \ ,\nonumber \\
\zeta_-(s,x,y)&=&i\ \sum_n \l_n^{-2s} \big(\p_n(x)\p_n^\dag(y)-\p^*_n(x)(\p_n^*)^\dag(y)\big)  \ .
\ea
Obviously, $\zeta^\dag_+(s,x,y)=\zeta_+(s,y,x)$ and  $\zeta^\dag_-(s,x,y)=-\zeta_-(s,y,x)$.

The convergence properties of these zeta-functions depend essentially on the large-$n$ behaviour of the eigenvalues $\l_n^2$. The latter, in turn, is dictated by the leading 2-derivative term in $D^2$ which, by \eqref{spinorLapl2}, is the same as the one of the scalar Laplacian, and which is the same as in flat space. It follows that,
as usual,  these zeta-functions are convergent expressions for $\Re s>1$ and are otherwise defined by analytical continuation.

Denoting the Dirac trace by $\trD$, we have
\ba\label{Zetaint}
\int \d^2 x \sqrt{g}\, \trD\, \zeta_+(s,x,x)&=&\sum_n \l_n^{-2s} \int\d^2 x \sqrt{g}\, \big(\chi^\dag_n(x)\chi_n(x)+\f_n^\dag(x)\f_n(x)\big) =
2 \sum_n \l_n^{-2s} \equiv \zeta(s) \ ,\nonumber\\
\int \d^2 x \sqrt{g}\, \trD \zeta_-(s,x,x)&=&\sum_n \l_n^{-2s} \int\d^2 x \sqrt{g}\, \big(\f^\dag_n(x)\chi_n(x)-\chi_n^\dag(x)\f_n(x)\big)=0 \ .
\ea
Note (again) that $2 \sum_n \l_n^{-2s}$ can be interpreted as $\sum \L_n^{-s}$ where $\L_n$ is either $\l_n^2$ or $(-\l_n)^2$, so that $\zeta(s)$ is actually the zeta function of $D^2$.

%%%%%%%%%%%%%%%%%%%%%%%%%%%%%%%%%%%%%%%%%%%%%%%%%%%%%%%%%
\subsubsection{Local heat kernels}

We similarly define local heat kernels that are again $2\times 2$-matrices :
\ba\label{Kdefs}
K_+(t,x,y)&=&\sum_n e^{-\l_n^2 t} \big(\chi_n(x)\chi_n^\dag(y)+\f_n(x)\f_n^\dag(y)\big)
= \sum_n e^{-\l_n^2 t} \big(\p_n(x)\p_n^\dag(y)+\p^*_n(x)(\p_n^*)^\dag(y)\big)  \ ,\nonumber \\
K_-(t,x,y)&=&\sum_n e^{-\l_n^2 t}  \big(\chi_n(x)\f_n^\dag(y)-\f_n(x)\chi_n^\dag(y)\big) 
= i\ \sum_n  e^{-\l_n^2 t}  \big(\p_n(x)\p_n^\dag(y)-\p^*_n(x)(\p_n^*)^\dag(y)\big)\ , \nonumber \\ 
\ea
as well as
\be\label{KKdef}
{\cal K}(t,x,y)=\frac{1}{2} K_+(t,x,y) - \frac{i}{2} K_- (t,x,y) = \sum_n e^{-\l_n^2 t} \p_n(x)\p_n^\dag(y) \ .
\ee
Note that $K_\pm$ are real functions and thus constitute the real and imaginary parts of ${\cal K}$.
All three, $K_+$, $K_-$ and ${\cal K}$ satisfy 
\be\label{heateq}
\left(\frac{\d}{\d t} + D_x^2\right) K_{\pm}(t,x,y) = 0 \ ,
\ee
where the subscript $x$ on $D^2$ indicates that the derivatives are with respect to $x$.

It is important to note that $K_+$ contains the full sum of all eigenfunctions of $D$, namely the $\p_n$ and the $\p_n^*$. Thus if one used a different basis for these eigenfunctions one would get the same $K_+$. Moreover, the $\p_n$ and the $\p_n^*$ appear symmetrically. We can write $K_+(t,x,y)=\bra{x} \sum_{\l_n^2} e^{-\l_n^2 t} P(\l_n^2) \ket{y}$ where $P(\l_n^2)$ is the projector on the eigenspace of $D^2$ with eigenvalue $\l_n^2$. This makes clear that one could use any basis of eigenfunctions of $D^2$. This also means that one should be able to obtain $K_+$  uniquely by solving the heat equation \eqref{heateq} for the operator $D^2$ with the appropriately prescribed short-distance singularity. However, this is not true for the imaginary part $K_-$ of ${\cal K}$. We see from \eqref{Kdefs} that the definition of $K_-$ is not simply a sum over all eigenfunctions of $D$, but that we made a certain distinction between the eigenfunctions with $\l_n>0$ and those with $\l_n<0$. Clearly, the operator $D^2$ does not make this distinction, and hence, one cannot simply get the $K_-$ by solving \eqref{heateq}. However, one can write $K_-$ in terms of an auxiliary quantity we call $L(t,x,y)$ as
\be\label{K-Lrel}
K_-(t,x,y)=D_x L(t,x,y) \ ,
\ee
where
\ba\label{Ldef}
L(t,x,y)&=&i \sum_n \frac{e^{-\l_n^2 t} }{\l_n} \big(\chi_n(x)\chi_n^\dag(y)+\f_n(x)\f_n^\dag(y)\big) 
=  i \sum_n \frac{e^{-\L_n t} }{\sqrt{\L_n}}   \big(\p_n(x)\p_n^\dag(y)+\p^*_n(x)(\p_n^*)^\dag(y)\big) \nonumber\\
&=& 2 i \, \Re\, \sum_n \frac{e^{-\L_n t} }{\sqrt{\L_n}}   \p_n(x)\p_n^\dag(y)\ .
\ea
We see that $L$ can now be constructed from the eigenfunctions and eigenvalues of $D^2$ only.
Another property of $K_-$ concerns is matrix structure at coinciding points~:
\be\label{Kmxx}
K_-(t,x,x) \sim \g_* \ .
\ee 
Indeed, if we call $a_n$ and $b_n$ the two complex components of $\p_n(x)$, then $\p_n(x)\p_n^\dag(x)-\p^*_n(x)(\p_n^*)^\dag(x)=2 \Im(b_n a_n^*)\, \g_*$.

The zeta-functions can be related to the heat kernels as usual by
\be\label{zetaK}
\zeta_\pm(s,x,y)= \frac{1}{\G(s)} \int_0^\infty \d t\, t^{s-1}\, K_\pm (t,x,y) \ .
\ee
Since $K$ vanishes exponentially for large $t$, any divergences of the integral occur from the region $t\to 0$. Thus any singularities (poles) of the zeta-functions are related to the small-$t$ behaviour of the heat kernel. Furthermore, since $\frac{1}{\G(s)}$ vanishes at $s=0,-1,-2,\ldots$, the finite values of the zeta-functions at $s=0,-1,-2,\ldots$ are also determined by the divergences of the integral due to the small-$t$ behaviour of $K$. In turn, for $K_+$ this small-$t$ behaviour can be determined in an asymptotic expansion from the differential equation \eqref{heateq}, with the leading behaviour being the same as in flat space, and the subleading terms being given in terms of the local curvature and derivatives of the curvature. Of particular interest will be the expansion at coinciding points $x=y$. In particular, one can show that\footnote{
A simple way to understand this result is to note that the heat kernel for the scalar Laplacian is $\frac{1}{4\pi t}\Big( 1 + \frac{\cR}{6}t+ \ldots)$, that the trace gives a factor 2, that the spinor Laplacian gives the same result to this order, and that the additional $\frac{\cR}{4}+m^2$ in $D^2$ simply gives an extra $e^{-(\cR/4+m^2)t}\simeq 1 -(\cR/4 + m^2)t$, to this order.}
\be\label{K+shortdist}
\trD K_+(t,x,x)=\frac{1}{2\pi t}\Big( 1 -\big(\frac{\cR(x)}{12}+m^2\big) \, t + {\cal O}(t^2) \Big) \ .
\ee
This determines
\be\label{zetapluszero}
\trD \zeta_+(0,x,x)=-\frac{1}{2\pi}  \big(\frac{\cR(x)}{12}+m^2\big) \ .
\ee
As far as the matrix structure of $\zeta_-(s,x,x)$ is concerned, it follows from \eqref{Kmxx} that 
\be\label{zetamooinsxx}
\zeta_-(s,x,x)\sim \g_* \ .
\ee

%%%%%%%%%%%%%%%%%%%%%%%%%%%%%%%%%%%%%%%%%%%%%%%%%%%%%%%%%
\subsubsection{Green's functions}\label{Green}

The Green's function $S(x,y)$ of the Dirac operator $D$ is a $2\times 2$-matrix solution of
\be\label{Sdef}
D_x S(x,y) = \frac{\dd(x-y)}{\sqrt{g}}\ {\bf 1}_{2\times 2} \ ,
\ee
while we denote $G$ the (also $2\times 2$-matrix) Green's function of $D^2$:
\be\label{Gdef}
D^2_x G(x,y) =\frac{\dd(x-y)}{\sqrt{g}}\ {\bf 1}_{2\times 2} \ .
\ee
In terms of the eigenfunctions and eigenvalues we have
\ba\label{SGeigen}
S(x,y)&=-i&\sum_n \frac{1}{\l_n} \big(\chi_n(x)\f_n^\dag(y)-\f_n(x)\chi_n^\dag(y)\big) \ , \nonumber\\
G(x,y)&=&\sum_n \frac{1}{\l_n^2} \big(\chi_n(x)\chi_n^\dag(y)+\f_n(x)\f_n^\dag(y)\big) \ .
\ea
They are indeed solutions of \eqref{Sdef}, resp. \eqref{Gdef} as one sees using the completeness relation \eqref{completeness}.
It trivially follows from either \eqref{Sdef} and \eqref{Gdef}, or from \eqref{SGeigen},  that
\be\label{SGrel}
S(x,y)=D_x G(x,y) \ .
\ee
Comparing with \eqref{Zetadefs}, one sees that\footnote{
Note that this is consistent with the relations $S=DG$ \eqref{SGrel} and $K_-=DL$ \eqref{K-Lrel}. Indeed,  
\ba
S(x,y)&\hskip-3.mm=\hskip-3.mm&-i\zeta_-(\frac{1}{2},x,y)=\frac{-i}{\G(\frac{1}{2})}\int_o^\infty \d t t^{-1/2} K_-(t,x,y) = \frac{-i}{\G(\frac{1}{2})}  \int_o^\infty \d t t^{-1/2} \, D_x L(t,x,y) \nonumber\\
&\hskip-3.mm=\hskip-3.mm&D_x \frac{1}{\G(\frac{1}{2})}\sum_n  \int_o^\infty \d t t^{-1/2} \frac{e^{-\L_n t} }{\sqrt{\L_n}}   \big(\p_n(x)\p_n^\dag(y)+\p^*_n(x)(\p_n^*)^\dag(y)\big) 
=D_x \sum_n \frac{1}{\L_n}   \big(\p_n(x)\p_n^\dag(y)+\p^*_n(x)(\p_n^*)^\dag(y)\big)  \nonumber\\
&\hskip-3.mm=\hskip-3.mm&D_x G(x,y) \nonumber \ .
\ea
}
\be\label{SGzeta}
S(x,y)=-i\,\zeta_-(\frac{1}{2},x,y) \quad , \quad G(x,y)=\zeta_+(1,x,y) \ .
\ee
It follows from the orthonormality of the $\chi_n$ and $\f_n$ that
\be\label{SSGrel}
\int \d^2 z\sqrt{g(z)} S(x,z) S(z,y) = G(x,y) \ .
\ee
(This also follows from \eqref{SGrel} and \eqref{Sdef} upon integrating by parts.)

As is clear from \eqref{Dsquared} and \eqref{spinorLapl2}, the matrix structure of $D^2$ is ${\bf 1} (\ldots) + \g_* (\ldots)$, implying that $G$ must have the same structure\footnote{
Note that $G$ is related to $\zeta_+$ and $K_+$ for which these arguments are correct, contrary to what was the case for $K_-$.
}~:
\be\label{matrixstrG}
G(x,y)= G_0(x,y) \, {\bf 1}+ G_*(x,y) \, \g_*\ .
\ee

The matrix structure of $S$ is somewhat less trivial, in particular for non-vanishing mass. We have 
\be\label{matrixstrSD}
D=i\s_x \cD_1+i \s_z \cD_2+ m \s_y \ ,
\ee
(with $\cD_1=E_1^\m\del_\m -\frac{1}{4} E_2^\m\o_\m$ and $\cD_2=E_2^\m\del_\m +\frac{1}{4} E_1^\m\o_\m$).
We write
\be\label{matrixstrS}
S=\s_x S_1 + \s_z S_2 +\s_y S_* + {\bf 1} S_0 \ .
\ee
Then \eqref{Sdef} yields a system of 4 equations:
\ba
\cD_1 S_2-\cD_2 S_1 + m S_0=0  \quad &,& \quad i\cD_1 S_1 + i \cD_2 S_2 +m S_* = \frac{\dd}{\sqrt{g}}\ , \\
\cD_1 S_*-\cD_2 S_0 +im S_1 = 0 \quad &,& \quad \cD_2 S_* + \cD_1 S_0 + im S_2 = 0 \ .
\ea
For $m=0$ the equations  of the first line decouple from those of the second line. We may then set $S_*=S_0=0$, so that only $S_1$ and $S_2$ are non-vanishing.\footnote{
This  is consistent with the relation $S=D G$ which yields $S_*=m G_0$ and $S_0=m G_*$, as well as $S_1=i\cD_1 G_0 +\cD_2 G_*$ and $S_2=-\cD_1 G_* + i \cD_2 G_0$.} 
In particular, we have
\be\label{traceSm0}
{\rm for \ } m=0 \quad : \quad \trD S(x,y) = \trD \g_* S(x,y) = 0 \ ,
\ee
where $\trD$ denotes the Dirac trace.
Maybe more useful, for general $m$, is to combine \eqref{SGrel} and \eqref{matrixstrG} which yields
\be\label{matrixSstr2}
S(x,y)= (i\Nsl_x G_0(x,y)+mG_*(x,y)) + \g_* \big( -i\Nsl_x G_*(x,y)+mG_0(x,y)\big) \ .
\ee
It follows from this relation that
\be\label{StraceG}
\trD \g_* S(x,y) = m \trD G(x,y) \ .
\ee

%%%%%%%%%%%%%%%%%%%%%%%%%%%%%%%%%%%%%%%%%%%%%%%%%
\subsection{Perturbation theory}

We want to study how the eigenvalues $\l_n$ (or $\l_n^2$) change under conformal rescalings of the metric. The variation of the Dirac operator $D$ (or of $D^2$) has been obtained in sect.~\ref{confsec}, see eqs \eqref{Nslvar} or \eqref{D2confscal2}.

Under $D\to D+\dd D$ we have $\l_n\to \l_n+\dd \l_n$, as well as $\chi_n\to \chi_n+\dd\chi_n$ and $\f_n\to \f_n+\dd \f_n$:
\be\label{pert1}
(D+\dd D)(\chi_n+\dd\chi_n)=i(\l_n+\dd\l_n)(\f_n+\dd\f_n) \quad \Rightarrow \quad\
\dd D\chi_n + D \dd\chi_n = i \dd\l_n \f_n+i \l_n\dd\f_n\ .
\ee
Taking the inner product with $\f_n$ and using the hermiticity of $D$ one gets
\be\label{pert2}
\dd\l_n=-i(\f_n,\dd D\chi_n) + \l_n \Big( (\chi_n,\dd\chi_n)-(\f_n,\dd\f_n)\Big) \ .
\ee
Note that $\chi_n$ and $\f_n$ are normalised with the metric $g$, while $\chi_n+ \dd\chi_n$ and $\f_n+\dd\f_n$  are normalised with $e^{\dd\s} g$. This implies
\be\label{pertnorm}
(\chi_n,\dd\chi_n)=-\int\d^2 x \sqrt{g}\, \dd\s(x) \chi_n^\dag(x) \chi_n(x) \ ,
\ee
and similarly for $(\f_n,\dd\f_n)$. Using also $\dd D=i \dd\Nsl$,  can then rewrite \eqref{pert2} as
\be\label{pert3}
\dd\l_n=(\f_n,\dd \Nsl\chi_n) + \l_n \int \dd\s\big( \f_n^\dag \f_n-\chi_n^\dag \chi_n  \big) \ .
\ee
(Obviously $\int \ldots$ stands for $\int \d^2 x \sqrt{g} \ldots $.) Next, using \eqref{Nslvar} for $\dd\Nsl$, and integrating by parts,
\ba\label{perrt4}
(\f_n,\dd \Nsl\chi_n)&=&-\int\dd\s \f_n^\dag \Nsl\chi_n +\frac{1}{2}\int  \del_\l\dd\s \f_n^\dag \g^\l \chi_n
=-\int\dd\s\Big( \f_n^\dag \Nsl\chi_n +\frac{1}{2} (\Nsl\f_n)^\dag \chi_n +\frac{1}{2}\f_n\Nsl\chi_n\Big)\nonumber\\
&=&-\int\dd\s \Big( -i\f_n^\dag (D-m\g_*)\chi_n +\frac{i}{2} \big((D-m\g_*)\f_n\big)^\dag\chi_n -\frac{i}{2} \f_n^\dag (D-m\g_*)\chi_n\Big)\nonumber\\
&=&-\int\dd\s\Big(\frac{3}{2}\l_n\f_n^\dag\f_n -\frac{1}{2}\l_n\chi_n^\dag\chi_n +im\f_n^\dag\g_*\chi_n\Big) \ ,
\ea
so that finally
\be\label{pert5}
\dd\l_n=-\int \dd\s\Big( \frac{\l_n}{2} (\f_n^\dag\f_n+\chi_n^\dag\chi_n) + i m \f_n^\dag \g_* \chi_n\Big)\ .
\ee

This allows us to express the variation of $\zeta(s)$ as
\be\label{zetavar}
\dd\zeta(s)=2\sum_n \dd \l_n^{-2s}= -4s\sum_n \frac{\dd\l_n}{\l_n^{2s+1}}
=2s \int\dd\s\sum_n \Big( \frac{1}{\l_n^{2s}}  (\f_n^\dag\f_n+\chi_n^\dag\chi_n) + 2im \frac{1}{\l_n^{2s+1}} \f_n^\dag \g_*\chi_n \Big) \ .
\ee
Since $\f_n$, $\chi_n$ and $i\g_*$ are real, we have $\f_n^\dag i\g_* \chi_n=(\f_n^\dag i\g_* \chi_n)^\dag=\chi_n^\dag (-i) \g_* \f_n$, and we can rewrite $\dd\zeta(s)$ in terms of the local zeta-functions $\zeta_+(s,x,x)$ and $\zeta_-(s,x,x)$ as
\be\label{zetavar2}
\dd\zeta(s)=2s \int\dd\s  \Big( \trD \zeta_+(s,x,x) + im \trD \g_* \zeta_-(s+\frac{1}{2},x,x) \Big) \ .
\ee
For the derivative we obviously get
\ba\label{zetaprimevar}
\dd\zeta'(s)&=&2 \int\dd\s  \Big( \trD \zeta_+(s,x,x) + im \trD \g_* \zeta_-(s+\frac{1}{2},x,x) \Big)  \nonumber\\
&&+2 s\int\dd\s  \Big( \trD \zeta'_+(s,x,x) + m \trD i \g_* \zeta'_-(s+\frac{1}{2},x,x) \Big) \ .
\ea
We want to evaluate both $\zeta$ and $\zeta'$ at $s=0$. Now $\zeta_+(0,x,x)$ is regular (cf e.g. \eqref{zetapluszero}, and subsection \ref{singstr}) and\footnote{
If a meromorphic function is regular at a given point, then its derivative necessarily is also regular at this point.} so is $\zeta'_+(0,x,x)$.
Thus
\ba\label{zetaatzero}
\dd\zeta(0)&=&2m  \int\dd\s  \lim_{s\to 0} s \trD i  \g_* \zeta_-(s+\frac{1}{2},x,x)  \ , \\
\label{zetprimeatzero}
\dd\zeta'(0)&=&2 \int\dd\s  \trD \zeta_+(0,x,x) 
+2m \int\dd\s  \lim_{s\to 0} \Big( \trD i \g_* \zeta_-(s+\frac{1}{2},x,x) + s  \trD i \g_* \zeta'_-(s+\frac{1}{2},x,x) \Big) \ . \nonumber\\
\ea

%%%%%%%%%%%%%%%%%%%%%%%%%%%%%%%%%%%
\subsection{Singularity structure of the local zeta-functions and the Green's functions}\label{singstr}

As usual we first need to establish the small-$t$ (and hence also short-distance) behaviours of the heat kernels $K_\pm(t,x,y)$, from which the singularity structure of the $\zeta_\pm(s,x,x)$ can be deduced. Before trying to make general statements, it is useful to look at the very simple example off the flat torus discussed in sect. \ref{flattorus}.

%%%%%%%%%%%%%%%%%%%%%%%%%%%%%%%%%%%%%%%%%%%%%%%%%%%%%%%%%
\subsubsection{The flat torus}

From the explicit form of the normalised eigenfunctions $\p_{\vec{n}}$ of $D$ on the flat torus given in \eqref{eigentorussol} we find 
\ba\label{calKtorus}
{\cal K}(t,x,y)&=&\sum_{n_1,n_2} e^{-\l^2_{\vec{n}}t} \p_{\vec{n}}(x)\p_{\vec{n}}^\dag(y) \nonumber\\
&=&\frac{1}{8\pi^2} \sum_{n_1,n_2} e^{-(n_1^2+n_2^2+m^2)t} e^{in_1(x^1-y^1)+in_2(x^2-y^2)}
\frac{1}{\l_{\vec{n}}} \begin{pmatrix} \l_{\vec{n}}-n_2 & -n_1-im \\  -n_1+im & \l_{\vec{n}}+n_2 \end{pmatrix} \ .
\ea
Note that in the real part the terms odd under $n_1\to - n_1$ or $n_2\to - n_2$ drop out of the sum :
\ba\label{Kplustorus}
K_+(t,x,y)&=&{\cal K}(t,x,y)+{\cal K}^*(t,x,y) = \frac{1}{4\pi^2} \sum_{n_1,n_2} e^{-(n_1^2+n_2^2+m^2)t} e^{in_1(x^1-y^1)+in_2(x^2-y^2)}
 \ {\bf 1}_{2\times 2}  \nonumber\\
 &=&  \frac{1}{4\pi^2} \, e^{-m^2 t}\  \t_3\big( \n=\frac{x^1-y^1}{2\pi}\big\vert \tau=i\frac{t}{\pi}\big) \ 
 \t_3\big( \n=\frac{x^2-y^2}{2\pi}\big\vert \tau=i\frac{t}{\pi}\big)\  {\bf 1}_{2\times 2} \ .
\ea
The well-known modular transformation of the Jacobi theta-function $\t_3$ 
\be\label{modular}
\t_3(\n\vert\tau)=\frac{1}{\sqrt{-i\tau}}e^{-i\pi \n^2/\tau} \t_3\big( \frac{\nu}{\tau}\big\vert -\frac{1}{\tau}\big)
\ee
allows us to immediately get the small-$t$ behaviour of $K_+(t,x,y)$ as
\be\label{Kplusastorus}
K_+(t,x,y) = \frac{1}{4\pi t} \ \exp\Big( - \frac{(x^1-y^1)^2+(x^2-y^2)^2}{4t}\Big)\   e^{-m^2 t}\  {\bf 1}_{2\times 2} \ \Big( 1 + {\cal O}\big( e^{-\pi^2/t}\big) \Big) \ .
\ee
The leading piece coincides, of course, with the well-known answer on ${\bf R}^2$.

On the other hand, the imaginary part of ${\cal K}(t,x,y)$ contains pieces proportional to $\s_x$, $\s_z$ and $\g_*=\s_y$. As noted above, this imaginary part $K_-$ cannot be obtained from the knowledge of $D^2$, but one needs to know the eigenfunctions of $D$. However, as also noted, $K_-$ is given by $K_-(t,x,y)=D_x L(t,x,y)$ where the quantity $L$ defined in \eqref{Ldef} is much simpler. At present it is given by $2i$ times the real part of the sum \eqref{calKtorus} with an extra factor $\frac{1}{\l_{\vec{n}}}$ inserted. Then again the terms odd under $n_1\to  n_1$ or $n_2\to -n_2$ drop out of the sum and we get
\be\label{Ltorus}
L(t,x,y)=\frac{i}{4\pi^2}\, e^{-m^2 t}\,  {\cal L}(t,x-y) \ {\bf 1}_{2\times 2} \quad , \quad
 {\cal L}(t,z)=
\sum_{n_1,n_2} \frac{1}{\l_{\vec{n}}}e^{-(n_1^2+n_2^2)t} e^{in_1 z^1+in_2 z^2}\, 
 \ .
\ee
Below, we will explicitly evaluate this sum for small $t$.

Mostly, we only need the heat kernels at coinciding points, in which case again all terms that are odd under $n_1\to -n_1$ or $n_2\to -n_2$ drop out, in the real and in the imaginary parts~:
\be\label{calKtorusxx}
{\cal K}(t,x,x)=\frac{1}{8\pi^2} \sum_{n_1,n_2} e^{-(n_1^2+n_2^2+m^2)t}
\Big( {\bf 1} + \frac{m}{\l_{\vec{n}}} \g_* \Big)\ .
\ee
The sum multiplying $\sim {\bf 1}$ was just computed exactly and can be read from \eqref{Kplusastorus}. A more generic way to obtain the small-$t$ behaviour is by noting that it is determined by the large eigenvalues, and for large eigenvalues the sums can be replaced by integrals :
\be\label{sumintegral}
\frac{1}{8\pi^2} \sum_{n_1,n_2} e^{-(n_1^2+n_2^2+m^2)t}\simeq \frac{e^{-m^2 t} }{8\pi^2}\int \d n_1 \d n_2 \, e^{-(n_1^2+n_2^2)t}
=\frac{e^{-m^2 t} }{8\pi^2} \frac{\pi}{t}=\frac{e^{-m^2 t} }{8\pi t} \ ,
\ee
which correctly reproduces the leading small-$t$ behaviour. Similarly, for the sum multiplying $\g_*$ in \eqref{calKtorusxx} we have
\ba\label{sumintegral2}
&&\frac{1}{8\pi^2} \sum_{n_1,n_2} e^{-(n_1^2+n_2^2+m^2)t} \frac{m}{\l_{\vec{n}}} \simeq
\frac{m}{8\pi^2} \int\d n_1\d n_2\, \frac{e^{-(n_1^2+n_2^2+m^2)t}}{\sqrt{n_1^2+n_2^2+m^2}}
=\frac{m}{8\pi} \int_0^\infty \d \xi \frac{e^{-(\xi+m^2)t}}{\sqrt{\xi+m^2}} \nonumber\\
&&=\frac{m}{8\pi} \int_{m^2}^\infty \d \xi \frac{e^{-\xi t}}{\sqrt{\xi}}
=\frac{m}{8\pi\sqrt{t}} \int_{m^2 t}^\infty \d \xi \frac{e^{-\xi}}{\sqrt{\xi}}
=\frac{m}{4\pi\sqrt{t}} \int_{m \sqrt{t}}^\infty \d z e^{-z^2}
=\frac{m}{8\sqrt{\pi t\ }}  \big(1 +{\cal O}(m\sqrt{t})\big) \ . \hskip1.cm 
\ea
Thus, we find for the flat torus
\be\label{calKtorusxx2}
{\cal K}(t,x,x)\simeq \frac{1}{8\pi t}\ {\bf 1} + \frac{m}{8\sqrt{\pi t\ }} \g_* +\ldots \ ,
\ee
where the unwritten terms $+\ldots$ are finite as $t\to 0$. We identify
\be\label{Kpmtorus}
K_+(t,x,x)=  \frac{1}{4\pi t}\ {\bf 1}  +\ldots  \quad , \quad K_-(t,x,x) = \frac{m}{4\sqrt{\pi t\ }}\ i \g_* +\ldots  \ .
\ee
As discussed above, the small-$t$ behaviours of $K_\pm(t,x,x)$  translate into possible poles of the corresponding local zeta-functions $\zeta_\pm(s,x,x)$ :
\ba\label{toruszeta}
&&\zeta_+(s,x,x) \sim_{s\to 1} \ \frac{{\bf 1} }{4\pi (s-1)} + {\rm finite}\ 
\ , \quad
\zeta_+(0,x,x) = {\rm finite}\times {\bf 1} \ , \nonumber\\
&&\zeta_-(s,x,x) \sim_{s\to 1/2}\  \frac{m i \g_* }{4\pi (s-\frac{1}{2} )}  + {\rm finite} \ .
\ea

Let us come back to the evaluation of the small-$t$ asymptotics of $K_-(t,x,y)$. To do so we look at the small-$t$ asymptotics of $L(t,x,y)$. We let $x^i-y^i=z^i\equiv z_i$ and write $|z|=\sqrt{z_1^2+z_2^2}$. For small $t$, the sum ${\cal L}(t,z)$ is again dominated by the large eigenvalues and we replace the sum over $n_1, n_2$  by an integral :
\be\label{Lintegral}
{\cal L}(t,z)\simeq  \int \d n_1 \d n_2 \frac{1}{\l_{\vec{n}}} \, e^{-(n_1^2+n_2^2)t} e^{i n_1 z^1 + i n_2 z^2} \ .
%=e^{-|z|^2/(4t)}\ \int \d n_1 \d n_2  \frac{e^{-t (n_1-n_1^*)^2- t(n_2-n_2^*)^2} }{\l_{\vec{n}}} \ ,
\ee
%where $n_1^*=\frac{i}{2t} z^1$ and $n_2^*=\frac{i}{2t}z^2$ are the ``saddle-point values". Of course, $t$ is small, and we do not really have a well-defined saddle-point, so we cannot simply replace $\l_{\vec{n}}$ by $\l_{\vec{n}^*}$.
%For $m=0$ we can go further. First suppose that $|z|\ne 0$. We set $n_i=\a\, \xi_i/\sqrt{t}$ and let $\a$ become very large so that we have a well-defined saddle point. Then the integal in \eqref{Lintegral} is 
%\ba
%{\cal L}(t,z)&\simeq &e^{-|z|^2/(4t)}\ \frac{\a}{\sqrt{t}}\, \int \d\xi_1\d\xi_2\, 
%&=&\frac{\a^2}{t}\frac{e^{-|z|^2/(4t)}\ }{\sqrt{(n_1^*)^2+(n_2^*)^2}}\, \int \d\tilde\xi_1\d\tilde\xi_2\,  e^{-\a^2 \tilde\xi_2^2 -\a^2\tilde\xi_2^2}
%=\frac{\pi}{t}\frac{e^{-|z|^2/(4t)}\ }{\sqrt{(n_1^*)^2+(n_2^*)^2}}\, \nonumber\\
%&=&\pm i \frac{\pi}{\sqrt{t}}  \Big(\frac{4t}{|z|^2}\Big)^{1/2} e^{-|z|^2/(4t)}
%^=\pm i \frac{2 \pi}{|z|}  \, e^{-|z|^2/(4t)}
%\  \quad {\rm if}\ |z|\ne 0 \ .
%\ .
%\ea
First, if $|z|=0$, we get for small $t$ (cf \eqref{sumintegral})
\be\label{Lforz0}
{\cal L}(t,z=0)\simeq \int \d n_1 \d n_2\  \frac{e^{-t (n_1^2+n_2^2)} }{\sqrt{n_1^2+n_2^2}} 
=2\pi\int_0^\infty \d n \, n\, \frac{e^{-tn^2}}{n}=\pi \sqrt{\frac{\pi}{t}} \ .
\ee
More generally, we have in terms of the Bessel function of the first kind $J_0$
\ba\label{Lintbessel}
{\cal L}(t,z)&\simeq& \int_0^\infty \d n \frac{n}{\l_{\vec{n}}} e^{-n^2 t} \int_0^{2\pi} \d\t e^{i n |z| \cos\t}
= 2\pi \int_0^\infty \d n \frac{n}{\l_{\vec{n}}} e^{-n^2 t}  J_0(n |z|)\nonumber\\
 &=&\frac{2\pi}{\sqrt{t}} \int_0^\infty \d \xi \frac{\xi}{\sqrt{\xi^2+m^2 t}} \ e^{-\xi^2}  J_0\big(\xi \frac{|z|}{\sqrt{t}}\big)
\simeq \frac{2\pi}{\sqrt{t}} \int_0^\infty \d \xi  \ e^{-\xi^2}  J_0\big(\xi \frac{|z|}{\sqrt{t}}\big) \ .
\ea
For $z=0$ one just gets back $\pi \sqrt{\frac{\pi}{t}}$. So let us assume now that $z\ne 0$. The last integral can be found e.g. in Erdelyi et al. (Bateman manuscript project, higher transcendental function, vol 2, sect. 7.7.3, eq 23) so that
\be\label{Lintbessel2}
{\cal L}(t,z)\simeq  \frac{\pi^{3/2}}{\sqrt{t}} e^{-|z|^2/(8t)} I_0 \big( \frac{|z|^2}{8t}\big)  \quad , \quad |z|\ne 0  \ ,
\ee
where $I_0$ is the modified Bessel function of the first kind. Its asymptotic for large argument is $I_0(a)\sim \frac{e^a}{\sqrt{2\pi a}}$
so that finally
\be\label{Lintbessel3}
{\cal L}(t,z)\simeq   \frac{\pi^{3/2}}{\sqrt{t}} \sqrt{\frac{4t}{\pi |z|^2}} = \frac{2\pi}{|z|} \quad , \quad |z|\ne 0 \ .
\ee
Alternatively, one can perform the two integrations in the reverse order. Then (within the same approximation $\frac{n}{\l_{\vec{n}}}\simeq 1 + {\cal O}(m^2 t)$) we have
\ba\label{Lint3}
{\cal L}(t,z)&\simeq&\int_0^\infty \d n e^{-n^2 t} \int_0^{2\pi} \d\t\, e^{i n |z| \cos\t}
=  \int_0^{2\pi}\d \t  \int_0^\infty \d n\, e^{-n^2 t}e^{i n |z| \cos\t}\nonumber\\
&=&\frac{1}{2} \int_0^{2\pi}\d \t \, \int_{-\infty}^\infty \d n e^{-n^2 t}e^{i n |z| \cos\t}
=\frac{1}{2} \int_0^{2\pi}\d \t\,  e^{-|z|^2 \cos^2\t/(4t)} \int_{-\infty}^\infty \d n e^{-t (n-i |z|\cos\t/(2t))^2}\nonumber\\
&=&\frac{1}{2}\sqrt{\frac{\pi}{t}} \int_0^{2\pi}\d \t \, e^{-|z|^2 \cos^2\t/(4t)} 
=\frac{1}{2} \sqrt{\frac{\pi}{t}}  \int_0^{2\pi}\d \t \, e^{-|z|^2 \sin^2\t/(4t)}  \ .
\ea
This is valid whether $|z|$ vanishes or not.
If $|z|=0$, the $\t$-integral is simply $2\pi$ and one gets back \eqref{Lforz0}. If $|z|\ne 0$ and $t\to 0$, the $\t$-integral is dominated by the two saddle-points where $\sin\t$ vanishes:
\ba\label{Lint4}
{\cal L}(t,z)&\hskip-3.mm\simeq\hskip-3.mm& \sqrt{\frac{\pi}{t}}  \int_{-\pi/2}^{\pi/2}\d \t \, e^{-|z|^2 \t^2/(4t)} 
\simeq\sqrt{\frac{\pi}{t}}  \int_{-\infty}^\infty\d \t\,  e^{-|z|^2 \t^2/(4t)} 
= \sqrt{\frac{\pi}{t}} \sqrt{\frac{4 \pi t}{|z|^2}}
=\frac{2\pi}{|z|}  \ , \ |z|\ne 0 \ ,
\ea
in agreement with \eqref{Lintbessel3}. However, we need an expression that still allows us to take both limits, $|z|\to 0$ and $t\to 0$.
Note that the first expression in \eqref{Lint4} indeed still is valid wether $|z|=0$ or not. Thus
\be\label{Lallz}
{\cal L}(t,z)\simeq \sqrt{\frac{\pi}{t}}  \int_{-\pi/2}^{\pi/2}\d \t \, e^{-|z|^2 \t^2/(4t)}  \quad , \quad |z|=0\ {\rm or}\ |z|\ne 0 \ .
\ee
We conclude
\be\label{Ltorusfinal}
L(t,x,y)\simeq_{t\to 0} \frac{i}{4 \pi\sqrt{\pi t\,}} \, {\bf 1}_{2\times 2}\  \int_{-\pi/2}^{\pi/2}\d \t \, e^{-\ell^2(x,y) \,\t^2/(4t)} \ ,
\ee
where at present $\ell^2(x,y)=|x-y|^2$.

%%%%%%%%%%%%%%%%%%%%%%%%%%%%%%%%%%%%%%%%%%%%%%%%%%%%%%%%%
\subsubsection{General statements}

It must be possible to make general statements  about the leading small-$t$ behaviour of $K_+$ based on the general form of $D^2$, just as in the bosonic case where the leading term in the asymptotic expansion of $K(t,x,y)$ always is $\frac{1}{4\pi t}\exp{\big(-\ell^2(x,y)/(4t)\big)}$ due to the 2-derivative part of the Laplacian being always $g^{\m\n}\del_\m\del_\n$. (Here $\ell(x,y)$ is the geodesic distance between $x$ and $y$.) Indeed, as discussed above, we may obtain $K_+(t,x,y)$ solely from the differential equation \eqref{heateq}, but not $K_-(t,x,y)$.

For $K_+(t,x,y)$ we expect, just as for $G(x,y)$, a piece $\sim {\bf 1}$ and a piece $\sim \g_*$. The leading small-$t$ singularity will be contained in the piece $\sim {\bf 1}$ and is universal, so that:
\be\label{K+txygen}
\trD K_+(t,x,y) \sim_{t\to 0} \frac{1}{2\pi t}\, e^{-\ell^2(x,y)/(4t)} \ ,
\ee
as well as at coinciding points $x=y$
\be\label{Kstrexp}
\trD K_+(t,x,x)\sim_{t\to 0} \frac{1}{2\pi t}\big( 1 + a_1(x)t + a_2(x) t^2 + \ldots \big) \ .
\ee

For $K_-(t,x,y) = D_x L(t,x,y)$ we expect that the leading small-$t$ singularity of $L$ is again generic and hence given by \eqref{Ltorusfinal}, so that
\be\label{K-asymp}
K_-(t,x,y)\sim_{t\to 0} \  \frac{i}{4 \pi\sqrt{\pi t\,}}  \int_{-\pi/2}^{\pi/2}\d \t \, \big( i\Nsl_x + m\g_*\big)\, e^{-\ell^2(x,y) \,\t^2/(4t)} \ .
\ee
It follows that
\be\label{traceK-} 
\trD i\g_* K_-(t,x,y) \sim_{t\to 0} -\frac{m}{2\pi\sqrt{\pi t\, }} \int_{-\pi/2}^{\pi/2}\d \t \, e^{-\ell^2(x,y) \,\t^2/(4t)} \ ,
\ee
and at coinciding points
\be
\trD i\g_* K_-(t,x,x)\sim_{t\to 0}\  - \frac{m}{2\sqrt{\pi t\ }} \big( 1 + b_1(x)t + b_2(x) t^2 + \ldots \big) \ .
\ee

The zeta-functions $\zeta_\pm$ are then obtained by the integral transform \eqref{zetaK} and the small $t$-asymp\-totics of the $K_\pm$ translate into possible singularities of the $\zeta_\pm$. Below we will study the singularities of the $\zeta_\pm(s,x,y)$ for $x\ne y$. Here we just note that at coinciding points one has
\be\label{zetaexp}
\trD \zeta_+(s,x,x) = \frac{1}{2\pi (s-1)} + \trD \zeta_+^{\rm reg}(s,x,x) +C_+ +{\cal O}(s-1)\ ,
\ee
while $\tr_D i\g_* \zeta_-(s,x,x)$ has poles at $s=\frac{1}{2}, -\frac{1}{2}, -\frac{3}{2}, \ldots $. In particular, for $s\to \frac{1}{2}$ we have
\be\label{zeta-half}
\trD i\g_* \zeta_-(s,x,x) \sim_{s\to 1/2} \ - \frac{m}{2\pi (s-\frac{1}{2})} + \trD i\g_* \zeta^{\rm reg}_-(\frac{1}{2},x,x) +C_- \ ,
\ee
The exact values of the constants $C_\pm$ depend on the exact definitions of $\zeta_\pm^{\rm reg}$ given below.
It follows
\be\label{sonehalf}
\lim_{s\to 0}\Big( \tr_D i\g_* \zeta_-(s+\frac{1}{2},x,x) + s\, \tr_D i\g_* \zeta_-'(s+\frac{1}{2},x,x)\Big) = \tr_D i\g_* \zeta_-^{\rm reg}(\frac{1}{2},x,x)+C_- \ .
\ee

%%%%%%%%%%%%%%%%%%%%%%%%%%%%%%%%%%%%%%%%%%%%%%%%%%%%%%%%%
\subsubsection{Singularities of the Green's functions}

The short-distance singularity of the Green's function $G(x,y)$ is dictated by the term with the most derivatives in $D^2$, which is the $-g^{\m\n}\del_\m\del_\n$ in $-\D_{\rm spinor}$. Thus the short-distance singularity is the same as in the bosonic case, except for the additional identity matrix :
\be\label{Gshortsing}
G(x,y)\sim_{x\to y} -\frac{1}{4\pi} \ln \m^2\ell^2(x,y) \ {\bf 1}_{2\times 2} \ + \ {\rm regular} \ .
\ee

The fermionic Green's function $S(x,y)$ is related to $G(x,y)$ by $S(x,y)=D_x G(x,y)$ and it follows that the short-distance singularity of $S$ is given by
\be\label{Sshortsing}
S(x,y)\sim_{x\to y} -\frac{1}{4\pi}\, \big( i \Nsl_x + m\g_*\big) \ln \m^2\ell^2(x,y) \ {\bf 1}_{2\times 2} \ + \ {\rm regular} \ .
\ee
There is a leading singularity $\sim -\frac{i}{4\pi} \frac{\dslsmall\ell^2}{\ell^2}$ as well as subleading singularities $\sim \ln\mu^2\ell^2$.

%%%%%%%%%%%%%%%%%%%%%%%%%%%%%
\subsection{Renormalised Green's functions}

For the Green's function $G$ of $D^2$, we may define a regularized Green's function $G^{\rm reg}(x,y)$ by subtracting the short-distance singularity \eqref{Gshortsing}
\be\label{Greg}
G^{\rm reg}(x,y)= G(x,y)+ \frac{1}{4\pi} \ln \m^2\ell^2(x,y) \ {\bf 1}_{2\times 2}  \ .
\ee
The so-called renormalized Green's function at coinciding points $G_{\rm R}$  then is simply defined as 
\be\label{GR}
G_{\rm R}(y)=\lim_{x\to y} G^{\rm reg}(x,y) \ .
\ee
In complete analogy, we define $S^{\rm reg}(x,y)$ and $S_{\rm R}(y)$~:
\be\label{Sreg}
S^{\rm reg}(x,y)= S(x,y)+ \frac{1}{4\pi}\, \big( i \Nsl_x + m\g_*\big) \ln \m^2\ell^2(x,y) \ {\bf 1}_{2\times 2} 
\ee
and
\be\label{SR}
S_{\rm R}(y)=\lim_{x\to y} S^{\rm reg}(x,y) \ .
\ee
In particular, multiplying \eqref{Sreg} with $\g_*$ and taking the trace yields
\be\label{Sregtrace}
\trD \g_* S^{\rm reg}(x,y)= \trD \g_* S(x,y)+ \frac{m}{2\pi}\,  \ln \m^2\ell^2(x,y) \ .
\ee

We now want to study how these regularised Green's functions are related to regularised zeta-functions.
For $G$ the story is much the same as in the purely bosonic case. Since the leading small-$t$ behaviour of the heat-kernel $K_+(t,x,y)$ is $\frac{1}{4\pi t}e^{-\ell^2(x,y)/(4t)}\, {\bf 1}_{2\times 2}$, we define a regularised zeta-function as
\ba\label{zeta+reg}
\zeta_+^{\rm reg}(s,x,y)&\hskip-3.mm=\hskip-3.mm&\frac{1}{\G(s)}\int_{1/\m^2}^\infty \d t\, t^{s-1} K_+(t,x,y) 
=\zeta_+(s,x,y) -\frac{1}{\G(s)}\int_0^{1/\m^2} \d t\, t^{s-1} \frac{e^{-\ell^2(x,y)/(4t)}}{4\pi t} \, {\bf 1}_{2\times 2}\nonumber\\
&\hskip-3.mm=\hskip-3.mm&\zeta_+(s,x,y) -\frac{(\m^2)^{1-s}}{4\pi \G(s)} \int_1^\infty \frac{\d u}{u} \, u^{1-s} e^{-\m^2 u\, \ell^2(x,y)  /4} 
\, {\bf 1}_{2\times 2}\nonumber\\
&=& \zeta_+(s,x,y) -\frac{(\m^2)^{1-s}}{4\pi \G(s)} E_s\big(\frac{\m^2 \ell^2(x,y) }{ 4}\big)\,  {\bf 1}_{2\times 2}\ ,
%\nonumber\\
\ea
where $E_s$ is the exponential integral function defined as 
\be\label{expint}
E_s(x)=\int_1^\infty \d u\, u^{-s}e^{-x u}\ . 
\ee
Its asymptotic expansions are well known and, in particular, for $s=1$ and $x\to y$, i.e. $\ell(x,y)\to 0$, we have
\be\label{Es1}
E_1\Big(\frac{\m^2\ell^2}{4} \Big) \sim_{\ell\to 0} -\g -\ln \frac{\m^2\ell^2 }{4} +{\cal O}\big( m^2\ell^2 \big) \ ,
\ee
Thus,
possible singularities in \eqref{zeta+reg} can occur for $s\to 1$ and/or $x\to y$ and come from the region of the integral where $t\to 0$. Because we cut off this region, this $\zeta_+^{\rm reg}$ clearly must be free of singularities. Indeed,
if we first set $s=1$ and then let $x\to y$, using $E_1(z) = -\g -\ln z +{\cal O}(z)$, we find
\ba\label{zeta+reg1}
\zeta_+^{\rm reg}(1,x,y)&&\hskip-5.mm
= \zeta_+(1,x,y)-\frac{1}{4\pi} E_1\big(\frac{\m^2 \ell^2(x,y) }{4}\big)\,  {\bf 1}_{2\times 2}
=G(x,y) -\frac{1}{4\pi} E_1\big(\frac{\m^2 \ell^2(x,y) }{4}\big) \,  {\bf 1}_{2\times 2}\nonumber\\
&&\hskip-5.mm\sim_{x\to y} G(x,y) +\Big(\frac{1}{4\pi} \ln \frac{\m^2 \ell^2(x,y) }{4} + \frac{\g}{4\pi}\Big) \,  {\bf 1}_{2\times 2} +{\cal O}(\m^2\ell^2)\nonumber\\
&&=G_{\rm R}(y) + \frac{\g-\ln 4}{4\pi} \,  {\bf 1}_{2\times 2}+{\cal O}(\m^2\ell^2) \ ,
% \nonumber\\
\ea
i.e.
\be\label{zeta+regxx}
G_{\rm R}(x)=\zeta_+^{\rm reg}(1,x,x)- \frac{\g-\ln 4}{4\pi} \,  {\bf 1}_{2\times 2} \ .
\ee
On the other hand, if we first let $x=y$, we get instead (for $\Re s>1$)
\ba\label{zeta+reg2}
\hskip-1.cm \zeta_+^{\rm reg}(s,x,x)&=&
 \zeta_+(s,x,x) -\frac{(\m^2)^{1-s}}{4\pi \G(s)} \int_0^1 \frac{\d v}{v} v^{s-1} \,  {\bf 1}_{2\times 2}
=  \zeta_+(s,x,x) -\frac{(\m^2)^{1-s}}{4\pi \G(s)}\  \frac{{\bf 1}_{2\times 2}}{s-1} \nonumber\\
&\sim_{s\to 1}& \,   \zeta_+(s,x,x) - \frac{{\bf 1}_{2\times 2}}{(4\pi)\, (s-1)} - \frac{C_+}{2}\, {\bf 1}_{2\times 2} \ ,
\ea
where 
\be\label{C+}
C_+= \frac{1}{2\pi} \big( \g-  \ln\m^2 \big)
\ee
One defines $G_\zeta(x)$ as
\be\label{Gzetadef}
G_\zeta(x)=\lim_{s\to 1} \Big( (\m^2)^{s-1} \zeta_+(s,x,x)  - \frac{{\bf 1}_{2\times 2}}{(4\pi)\, (s-1)} \Big) \ ,
\ee
so that
\be\label{Gzeta+regxx}
G_\zeta(x)=\zeta_+^{\rm reg}(1,x,x)+\frac{\g}{4\pi} \, {\bf 1}_{2\times 2} \ .
\ee
Comparing \eqref{zeta+regxx} and \eqref{Gzeta+regxx} we find
\be\label{GzetaGR}
G_\zeta(x)=G_R(x)+\frac{\g-\ln 2}{2\pi} \,  {\bf 1}_{2\times 2} \ .
\ee

Similarly, the Green's function $S(x,y)$ of the Dirac operator $D$ equals $-i\zeta_-(\frac{1}{2},x,y)$. Thus, obviously the latter is singular as $x\to y$. On the other hand we may consider $\zeta_-(s,x,x)$ which is regular as long as $s\ne \frac{1}{2}$ but which has a pole at $s=\frac{1}{2}$. As for $G$ and $\zeta_+$, we now want to define a regularized $\zeta_-^{\rm reg}(s,x,y)$ by removing the singular part, so that it is regular at $s=\frac{1}{2}$ and $x=y$~:
\be\label{zeta-reg}
\zeta_-^{\rm reg}(s,x,y)=\frac{1}{\G(s)}\int_{1/\m^2}^\infty \d t\, t^{s-1} K_-(t,x,y) 
=\zeta_-(s,x,y) -\frac{1}{\G(s)}\int_0^{1/\m^2} \d t\, t^{s-1} k^{\rm small}_-(t,x,y)\ ,
\ee
where $k^{\rm small}_-(t,x,y)$ is meant to be just the leading small-$t$ asymptotics of $K_-(t,x,y)$.
Upon multiplying with $i\g_*$, taking the trace and using \eqref{traceK-}  we get
\ba\label{zeta-reg2}
\trD i\g_*\zeta_-^{\rm reg}(s,x,y)
&=&\trD i\g_* \zeta_-(s,x,y) +\frac{1}{\G(s)} \frac{m}{2 \pi^{3/2}}  \int_{-\pi/2}^{\pi/2} \d\t\, \int_0^{1/\m^2} \d t\, t^{s-3/2}\,  e^{-\ell^2\t^2/(4t)} \nonumber\\
&=&\trD i\g_* \zeta_-(s,x,y) +\frac{1}{\G(s)} \frac{m \m^{1-2s}}{2 \pi^{3/2}}  \int_{-\pi/2}^{\pi/2} \d\t\, \int_1^\infty \d u\, u^{-1/2-s}\,  e^{-\m^2\ell^2\t^2\,u/4} 
\nonumber\\
&=&\trD i\g_* \zeta_-(s,x,y) + \frac{m \m^{1-2s}}{2 \pi^{3/2} \G(s)}  \int_{-\pi/2}^{\pi/2}E_{s+1/2}\Big(\frac{\m^2\ell^2}{4}\, \t^2 \Big) \ ,
\ea
where $E_s$ was defined in \eqref{expint}. We can now set $s=\frac{1}{2}$ and let $x\to y$, i.e. $\ell(x,y)\to 0$. Using again the asymptotics \eqref{Es1}(with $\ell^2\to \ell^2 \t^2$), 
and  doing the $\t$-integral, we get (recall $\zeta_-(\frac{1}{2},x,y)=iS(x,y)$)
\be\label{zeta-reg3}
\trD i\g_*\zeta_-^{\rm reg}(\frac{1}{2},x,y)
=-\trD \g_* S(x,y) - \frac{m}{2\pi} \Big( \g + \ln\m^2\ell^2(x,y) +2\ln\frac{\pi}{4}-2\Big) \ .
\ee
If we now let $x\to y$ and use \eqref{Sregtrace} and \eqref{SR} we find
\be\label{zeta-reg4}
-\trD i\g_*\zeta_-^{\rm reg}(\frac{1}{2},x,x)
=\trD \g_* S_{\rm R}(x) + \frac{m}{2\pi} \Big( \g +2\ln\frac{\pi}{4}-2\Big) \ .
\ee
If instead in the first line of \eqref{zeta-reg2} we first let $x=y$ and then $s\to \frac{1}{2}$, we get\footnote{
We have $\G(s)=\G(\frac{1}{2})\Big(1+\p(\frac{1}{2})(s-\frac{1}{2})\Big)$ with $\p(\frac{1}{2})=-2\ln2-\g$. Thus $\frac{1}{\G(s)}=\frac{1}{\sqrt{\pi}}\Big( 1+\big(2\ln2+\g)(s-\frac{1}{2})\Big)$.
}
\ba\label{zeta-reg5}
\trD i\g_*\zeta_-^{\rm reg}(s,x,x)
&=&\trD i\g_* \zeta_-(s,x,x) +\frac{1}{\G(s)} \frac{m}{2 \pi^{1/2}} \frac{\m^{1-2s}}{s-\frac{1}{2}}\nonumber\\
&\sim_{s\to 1/2}&\ \trD i\g_* \zeta_-(s,x,x) + \frac{m}{2\pi (s-\frac{1}{2})}-C_-\ ,
\ea
where
\be\label{C-def}
C_-=\frac{m}{2\pi}\Big(\ln\m^2-\ln 4 -\g\Big)  \ .
\ee
In analogy with $G_\zeta(x)$ we could try to define $S_\zeta(x)$ in terms of $\zeta_-(s,x,x)$ by subtracting its pole at $s=\frac{1}{2}$.
Since we only have given the relations for the traces with $\g_*$ we restrict to this case:
\be\label{Szeta}
\trD \g_* S_\zeta(x)=\lim_{s\to\frac{1}{2}}\Big( -(\m^2)^{s-1/2} \trD i \g_* \zeta_-(s,x,x) -\frac{m}{2\pi (s-\frac{1}{2})} \Big) \ ,
\ee
so that
\be\label{Szetazetareg}
\trD \g_* S_\zeta(x)= - \trD i\g_*\zeta_-^{\rm reg}(\frac{1}{2},x,x) +\frac{m}{2\pi} (\g+\ln4)\ .
\ee
Comparing \eqref{Szetazetareg} and \eqref{zeta-reg4} we conclude
\be\label{SzetaSR}
\trD\g_* S_\zeta(x)=\trD\g_* S_R(x)+\frac{m}{\pi} \Big( \g +\ln\frac{\pi}{2}-1\Big) \ .
\ee

%%%%%%%%%%%%%%%%%%%%%%%%%%%%%%%%%%%%%%%%%%%%%%%%
\section{The variation of the gravitational action}

%%%%%%%%%%%%%%%%%%%%%%%%%%%%%%%%%%%%%%%%%%%%%%%%%%%%%%%%%
\subsection{Expression in terms of $S_\zeta$ or $S_{\rm R}$}

Recall our formula \eqref{gravac4} for the gravitational action  in terms of $\zeta(0)$ and  $\zeta'(0)$. Their variations under infinitesimal conformal rescalings with $\dd\s(x)$ have been worked out  above and are given in \eqref{zetaatzero} and \eqref{zetprimeatzero}.
Combining \eqref{zetaatzero} with \eqref{zeta-half} we get (writing again explicitly the $\sqrt{g}=e^{2\s}\sqrt{\hat g}$ and using $A=\int\sqrt{g}$)
\be\label{deltazetazero2}
\dd\zeta(0)=-\frac{m^2}{\pi} \int\sqrt{g}\dd\s = -\frac{m^2}{2\pi} \dd \int\sqrt{g} =  -\frac{m^2}{2\pi} \dd A\ ,
\ee
while combining \eqref{zetprimeatzero} with \eqref{zetapluszero} and \eqref{sonehalf} we get
\be\label{deltazetazeroprime2}
\dd\zeta'(0)=-\int\sqrt{g}\dd\s \Big( \frac{\cR}{12\pi}+\frac{m^2}{\pi}\Big) + 2 m \int\sqrt{g} \dd\s \Big( \trD i\g_* \zeta_-^{\rm reg}(\frac{1}{2},x,x) +C_- \Big)\ .
\ee
Thus
\ba\label{sgravvar}
\dd S_{\rm grav}&=&\frac{1}{4}\dd\zeta'(0)+\frac{1}{4}\ln \m^2\, \dd\zeta(0) \nonumber\\
&=& -\frac{m^2}{8\pi} \big(1+\g+2\ln 2\big) \dd A    -\frac{1}{48\pi} \int \sqrt{g}\dd\s \cR 
+\frac{m}{2} \int\sqrt{g}\dd\s  \trD i\g_* \zeta_-^{\rm reg}(\frac{1}{2},x,x) \ .
\ea
Now, the terms $\sim \dd A$ are just the variation of a cosmological constant term and
\be\label{Liouville}
\int\sqrt{g}\dd\s \cR = \dd S_{\rm Liouville}
\ee
is the variation of the Liouville action. Note that the coefficient of $\dd S_{\rm Liouville}$ in $\dd S_{\rm grav}$  is exactly $\frac{1}{2}$ the one that occurs in the gravitational action of a bosonic scalar matter field. It is interesting to trace back how this coefficient occurs. For the bosonic scalar field and scalar Laplacian, the heat kernel coefficient $a_1$ is $\frac{\cR}{6}$. At present, $a_1$ is $\frac{\cR}{6}-\frac{\cR}{4}=-\frac{\cR}{12}$. But because we are dealing with fermions, the gravitational action has an overall minus sign, so that in the end one gets a $+\frac{\cR}{12}$, i.e. exactly one half the bosonic result.\footnote{Of course, we also had the $\frac{1}{4}$ multiplying the $\dd\zeta'(0)$ and $\dd\zeta(0)$ instead of a $\frac{1}{2}$ in the bosonic case. But this extra $\frac{1}{2}$ is offset by a factor $2$ coming from the Dirac traces.}
We conclude that
\be\label{sgravvar2}
\dd\Big( S_{\rm grav} +\frac{1}{48\pi} S_{\rm Liouville} +\frac{m^2}{8\pi} \big(1+\g+2\ln 2\big) A \Big)
= \frac{m}{2} \int\sqrt{g}\,\dd\s  \trD i\g_* \zeta_-^{\rm reg}(\frac{1}{2},x,x) \ .
\ee
We may use \eqref{Szetazetareg} to re-express this variation as
\be\label{sgravvar3}
\dd\Big( S_{\rm grav} +\frac{1}{48\pi} S_{\rm Liouville} +\frac{m^2}{8\pi} A \Big)
= -\frac{m}{2} \int\sqrt{g}\,\dd\s  \trD \g_* S_\zeta(x) \ ,
\ee
or instead using \eqref{zeta-reg4} as
\be\label{sgravvar4}
\dd\Big( S_{\rm grav} +\frac{1}{48\pi} S_{\rm Liouville} +\frac{m^2}{8\pi} \big(2\g+ 2 \ln\frac{\pi}{2}-1\big) A \Big)
= -\frac{m}{2} \int\sqrt{g}\,\dd\s  \trD \g_* S_{\rm R}(x) \ ,
\ee
It remains to characterise the right hand side of any of these last three equations and express it as a total variation of some appropriate quantity. We will focus on the last form which contains $S_{\rm R}$. 

%%%%%%%%%%%%%%%%%%%%%%%%%%%%%%%%%%%%%%%%%%%
\subsection{Conformal variations of the Green's function $S(x,y)$ and $S_{\rm R}(x)$}

Recall that $S(x,y)$ is the solution of $D_x S(x,y)\equiv (i\Nsl_x+m\g_*)S(x,y)=\dd(x-y)/\sqrt{g}$, see eq. \eqref{Sdef}. We want to determine  the variation of $S$ under a conformal rescaling. Consider two metrics $g$ and $\hat g$ related by $g=e^{2\s}\hat g$, cf \eqref{confresc}. Then, of course $\sqrt{g}=e^{2\s}\sqrt{\hat g}$ and from \eqref{Nslconf},  $\Nsl =e^{-3\s/2}\,\hat\Nsl \ e^{\s/2}$. Thus
\be\label{SandShatsol}
(i\hat\Nsl_x+m\g_*)S_{\hat g}(x,y)=\frac{\dd(x-y)}{\sqrt{\hat g}}
\quad , \quad
\big( i e^{-3\s(x)/2}\,\hat\Nsl \ e^{\s(x)/2} + m \g_* \big) S_{g}(x,y) =\frac{\dd(x-y)}{\sqrt{\hat g}} \, e^{-3\s(x)/2}e^{-\s(y)/2}\ .
\ee
We see that, for $m=0$, one simply has $S_g(x,y)=e^{-\s(x)/2} S_{\hat g}(x,y) e^{-\s(y)/2}$. This motivates us to define
\be\label{Stilde}
\St_g(x,y)=e^{\s(x)/2} S_{g}(x,y) e^{\s(y)/2} \ ,
\ee
so that the statement for $m=0$ simply is $\St_g=\St_{\hat g}$.  For non-vanishing $m$ we may rewrite the second equation
\eqref{SandShatsol} in terms of $\St$ as
\be
i\hat\Nsl \,\St_g(x,y)+ m\g_*  e^{\s(x)} \St_g(x,y) = \frac{\dd(x-y)}{\sqrt{\hat g}}  \ .
\ee
We let $\s=\dd\s$ be infinitesimal and let $\St_g=\St_{\hat g}+\dd\St$ so that $(i\hat\Nsl+m\g_*) \dd\St(x,y)=-m\g_* \dd\s(x) \St_{\hat g}(x,y)$. But we may now consider $\hat g$ as $g$ and simply rewrite this as (relabelling $x\to z$)
\be\label{deltaSt1}
D_z\,  \dd \St(z,y) \equiv (i\Nsl+m\g_*) \dd\St(z,y)=-m\g_* \dd\s(z) \St(z,y)=-m\g_* \dd\s(z) S(z,y) \ ,
\ee
where in the last step we used that $\dd\s$ is infinitesimal.
We then multiply  with  $S(x,z) \sqrt{g(z)}$  from the left,  integrate over $z$ and get, using the hermiticity of $D_z$,
\be\label{deltaST2}
\dd\St(x,y)=-m\int\d^2 z \sqrt{g}\, S(x,z)\g_* \dd\s(z) S(z,y) \ ,
\ee
or
\be\label{deltaST3}
\dd S(x,y) + \frac{1}{2} \big( \dd\s(x)+\dd\s(y)\big) S(x,y)=-m\int\d^2 z \sqrt{g}\, S(x,z)\g_* \dd\s(z) S(z,y) \ ,
\ee

Since $S_{\rm R}$ is obtained from $S$ by subtracting the short-distance singularity $S^{\rm sing}$, we need the conformal transformation of the latter:
\ba\label{deltaSsing}
\dd S^{\rm sing}(x,y) &=& \dd\Big(-\frac{1}{4\pi}\, \big( i \Nsl_x + m\g_*\big) \ln \m^2\ell^2(x,y) \ {\bf 1}_{2\times 2}\Big)
\nonumber\\
&=&-\frac{1}{4\pi} \Big(\, i\dd\Nsl_x \ln \m^2\ell^2(x,y)+  \big( i \Nsl_x + m\g_*\big) \frac{\dd\ell^2(x,y)}{\ell^2(x,y)} \Big) {\bf 1}_{2\times 2} \ .
\ea
We will only need the trace of this expression multiplied by $\g_*$, so that only the term $\sim m\g_*$ survives:
\be\label{deltaSsing2}
\trD \g_* \dd S^{\rm sing}(x,y)= - \frac{m}{2\pi} \frac{\dd\ell^2(x,y)}{\ell^2(x,y)} =  - \frac{m}{2\pi} \big( \dd\s(x)+\dd\s(y) \big) \ .
\ee
Obviously then,
\be\label{deltaSsing3}
\trD \g_* \Big(\dd S^{\rm sing}(x,y)  + \frac{1}{2} \big( \dd\s(x)+\dd\s(y)\big) S^{\rm sing}(x,y)\Big)
=   - \frac{m}{4\pi}\Big( 2 +  \ln \m^2\ell^2(x,y) \Big) \big( \dd\s(x)+\dd\s(y) \big) \ .
\ee
Subtracting  \eqref{deltaSsing3} from the trace with $\g_*$ of \eqref{deltaST3} we get
\ba\label{deltaSTreg}
\hskip-8.mm\trD \g_*\Big(\dd S^{\rm reg}(x,y) + \frac{1}{2} \big( \dd\s(x)+\dd\s(y)\big) S^{\rm reg}(x,y)\Big)
&\hskip-3.mm=\hskip-3.mm&-m\int\d^2 z \sqrt{g}\, \trD\g_* S(x,z)\g_* \dd\s(z) S(z,y) \nonumber\\
&&\hskip-3.mm+\ \frac{m}{4\pi}\Big( 2 +  \ln \m^2\ell^2(x,y) \Big) \big( \dd\s(x)+\dd\s(y) \big) \ .
\ea
One can check that the singularity of the integral in the first line is logarithmic as $x\to y$, and it is exactly cancelled by the one of the second line. In terms of $S_{\rm R}$ this can be restated as
\be\label{deltaSR}
e^{-\s(y)}\, \dd \trD \g_*\Big(e^{\s(y)}\, S_{\rm R}(y) \Big)
=\trD \g_*\Big(\dd S_{\rm R}(y) + \dd\s(y)S_{\rm R}(y)\Big)
=\ \frac{m}{\pi} \dd\s(y)
 -m F[y,\dd\s] \ ,  
 \ee
where we defined
\be\label{Fdef}
F[y,\dd\s] = \lim_{x\to y} \Big(\int\d^2 z \sqrt{g}\,  \dd\s(z) \, \trD\g_* S(x,z)\g_* S(z,y) -\ \frac{\dd\s(y)}{2\pi} \ln \m^2\ell^2(x,y) \Big)
\ .
\ee
We may  then rewrite the integral on the right-hand side of \eqref{sgravvar4} as
\be\label{sgravvar5}
\int\sqrt{g}\,\dd\s  \trD \g_* S_{\rm R} = \int \sqrt{\hat g}\ \dd\big( e^\s\big) \, e^\s  \trD \g_* S_{\rm R}
= \dd  \int \sqrt{g}\, \trD \g_* S_{\rm R} - \int \sqrt{g}\, e^{-\s} \,  \dd\big(e^\s  \trD \g_* S_{\rm R}\big) \ ,
\ee
where the integrand in last term is then given by \eqref{deltaSR}. Thus \eqref{sgravvar4}  becomes
\be\label{sgravvar6}
\dd\Big( S_{\rm grav} +\frac{1}{48\pi} S_{\rm Liouville} +\frac{m^2}{8\pi} \big(2\g+ 2 \ln\frac{\pi}{2}-3\big) A +\frac{m}{2}  \int \sqrt{g}\, \trD \g_* S_{\rm R} \Big)
= -\frac{m^2}{2} \int\sqrt{g}\, F[y,\dd\s]\ .
\ee
Recall from \eqref{StraceG} that $\trD\g_* S(x,y)=m \trD G(x,y)$. This same relation obviously is also valid for the corresponding singular parts that need to be subtracted to get $S^{\rm reg}$ and $G^{\rm reg}$. Hence also  $\trD\g_* S^{\rm reg}(x,y)=m \trD G^{\rm reg}(x,y)$, and in the coincidence limit~:
\be\label{SRGRrel}
\trD \g_* S_{\rm R} =m \trD G_{\rm R} \ .
\ee
We may thus rewrite \eqref{sgravvar6} as
\be\label{sgravvar6-bis}
\dd\Big( S_{\rm grav} +\frac{1}{48\pi} S_{\rm Liouville} +\frac{m^2}{8\pi} \big(2\g+ 2 \ln\frac{\pi}{2}-3\big) A +\frac{m^2}{2}  \int \sqrt{g}\, \trD G_{\rm R} \Big)
= -\frac{m^2}{2} \int\sqrt{g}\, F[y,\dd\s]\ .
\ee

Let us then study the properties of $F[y,\dd\s]$. Recall $S=(i\Nsl + m \g_*) G$ and, from \eqref{matrixstrG}, $G=G_0\, {\bf 1} + G_* \g_*$. It follows that $\g_* S \g_*=(-i\Nsl + m \g_*)G$, and taking also into account 
$\trD \g_*=\trD \g_* \Nsl =\trD \Nsl=0$, we get
\ba\label{TrgSgSstr}
\trD\g_* S(x,z)\g_* S(z,y)&=&-\trD i\Nsl_x G(x,z)\, i\Nsl_z G(z,y) + m^2 \trD G(x,z)G(z,y)\nonumber\\
&=& -\trD (i\Nsl_x+m\g_*) G(x,z)\, (i\Nsl_z+m\g_*) G(z,y) + 2 m^2 \trD G(x,z)G(z,y)\nonumber\\
&=& -\trD S(x,z)S(z,y) + 2 m^2 \trD G(x,z)G(z,y)\ ,
\ea
so that we can rewrite
 \ba\label{Frewrite}
F[y,\dd\s] &=&- \lim_{x\to y} \Big(\int\d^2 z \sqrt{g}\,  \dd\s(z) \, \trD S(x,z)S(z,y) +\ \frac{\dd\s(y)}{2\pi} \ln \m^2\ell^2(x,y) \Big)\nonumber\\
&+& 2 m^2 \int\d^2 z \sqrt{g}\,  \dd\s(z) \, \trD G(y,z)G(z,y)\ ,
\ea
where the integral of the last line is of course regular, so that we could set $x=y$. It will be useful to define a similar quantity without the $\dd\s$ as
 \be\label{Hdef}
H[y] =- \lim_{x\to y} \Big(\int\d^2 z \sqrt{g}\, \trD S(x,z)S(z,y) +\ \frac{1}{2\pi} \ln \m^2\ell^2(x,y) \Big)
+2 m^2 \int\d^2 z \sqrt{g}\,   \trD G(y,z)G(z,y)\ .
\ee
Using \eqref{SSGrel} we see that the term inside the bracket is just $\trD G^{\rm reg}(x,y)$ and thus
 \be\label{H2}
H[y] =-\trD G_{\rm R}(y)+ 2 m^2 \trD G_2(y,y) \ ,
\ee
where $G_2(x,y)=\zeta_+(2,x,y)$.
We want to express $F$ in terms of the variation of $H$, to leading order in a small $m$-expansion.

%%%%%%%%%%%%%%%%%%%%%%%%%%%%%%%%%%%%%
\subsection{The gravitational action on the sphere}

In this subsection, we will  restrict ourselves to the case where $i\Nsl$ has no zero-modes, so that all eigenvalues $\l_n$ are strictly positive, even for $m=0$.  This is the case, in particular, for spherical topology, but also  for the torus with a spin structure involving at least one anti-periodic boundary condition. The important thing is that in all these cases there is no zero-mode piece in the Green's functions one would need to remove in the $m\to 0$ limit and, hence, in a small mass expansion, we may consider all Green's function to be of order $m^0$. This would not be the case in the presence of zero-modes of $i\Nsl$ where $S$ would have a piece $\sim\frac{1}{m}$ and $G$ a piece $\sim\frac{1}{m^2}$. Thus in our previous equation for $F$ and $H$ we may already drop the terms $\sim m^2 GG$, and we rewrite
\be\label{Hrewrite}
e^{\s(y)} H(y)
=- \lim_{x\to y} \Big(\int\d^2 z \sqrt{\hat g}\, e^{\s(z)}\, \trD \St(x,z)\St(z,y) +\ \frac{e^{\s(y)}}{2\pi} \ln \m^2\ell^2(x,y) \Big)
+ {\cal O}(m^2) \ ,
\ee
where $\St$ defined in \eqref{Stilde} was such that $\dd\St={\cal O}(m)$, cf \eqref{deltaST2}. It follows immediately that
\ba\label{Hvar}
\dd\Big(e^{\s(y)} H(y)\Big)
&=&- \lim_{x\to y} \Big(\int\d^2 z \sqrt{\hat g}\, e^{\s(z)}\, \dd\s(z)\, \trD \St(x,z)\St(z,y) +\ \frac{e^{\s(y)}}{2\pi}\, \dd\s(y)\,  \ln \m^2\ell^2(x,y)\Big) \nonumber\\
&&-\ \frac{e^{\s(y)}}{\pi}\, \dd\s(y) + {\cal O}(m)\nonumber\\
&=&- e^{\s(y)}\lim_{x\to y} \Big(\int\d^2 z \sqrt{g}\, \dd\s(z)\, \trD S(x,z)S(z,y) +\ \frac{\dd\s(y)}{2\pi}\, \,  \ln \m^2\ell^2(x,y)\Big) 
 \nonumber\\
&&-\ \frac{e^{\s(y)}}{\pi}\, \dd\s(y) + {\cal O}(m)\nonumber\\
&=& e^{\s(y)} \Big( F[y,\dd\s] -\frac{\dd\s(y)}{\pi} \Big) + {\cal O}(m) \ .
\ea
Thus
\ba\label{FHmanip}
\int \sqrt{g} F +{\cal O}(m)&=&\int\sqrt{\hat g}\, e^\s \dd\big( e^\s H\big) + \frac{\dd A}{2\pi}
=\int\sqrt{\hat g}\, e^{2\s}\big( \dd\s H + \dd H\big)  + \frac{\dd A}{2\pi}\nonumber\\
&=&\frac{1}{2} \int\sqrt{\hat g}\, \Big(\dd\big( e^{2\s} H\big) + e^{2\s}\dd H \Big)+ \frac{\dd A}{2\pi}
= \frac{1}{2} \, \dd\, \int\sqrt{g} H + \frac{1}{2} \int\sqrt{g}\, \dd H + \frac{\dd A}{2\pi} \ .\nonumber\\
\ea
Now to lowest order in $m$, by \eqref{H2} we have simply $H=-\trD G_{\rm R}$ and thus
\be\label{Fexpr}
\int \sqrt{g} F = -  \frac{1}{2} \, \dd\, \int\sqrt{g}\, \trD G_{\rm R} - \frac{1}{2} \, \int\sqrt{g}\, \dd\, \trD G_{\rm R}
+ \frac{\dd A}{2\pi} +{\cal O}(m)\ .
\ee
Inserting this into eq. \eqref{sgravvar6-bis} for the gravitational action we get
\be\label{sgravvar8}
\dd\Big( S_{\rm grav} +\frac{1}{48\pi} S_{\rm Liouville} +\frac{m^2}{8\pi} \big(2\g+ 2 \ln\frac{\pi}{2}-1\big) A +\frac{m^2}{4} \int\sqrt{g}\, \trD G_{\rm R}\Big)
= \frac{m^2}{4} \int\sqrt{g}\, \dd \trD G_{\rm R}  +{\cal O}(m^3)\ .
\ee
By \eqref{GzetaGR} we may replace $G_{\rm R}=G_\zeta +\frac{1}{2\pi} (\ln 2-\g) {\bf 1}_{2\times 2}$ so that
\be\label{sgravvar9}
\dd\Big( S_{\rm grav} +\frac{1}{48\pi} S_{\rm Liouville} +\frac{m^2}{8\pi} \big(2 \ln\pi-1\big) A +\frac{m^2}{4} \int\sqrt{g}\, \trD G_\zeta\Big)
= \frac{m^2}{4} \int\sqrt{g}\, \dd \trD G_\zeta  +{\cal O}(m^3)\ .
\ee

Finally, we need $\dd\trD G_\zeta=\dd\trD G_{\rm R}$, up to terms of order $m$. It follows from \eqref{GR} and \eqref{Greg} that
\be\label{GRvar}
\dd G_{\rm R}(y)=\lim_{x\to y} \Big( \dd G(x,y) +\frac{{\bf 1}_{2\times 2}}{4\pi} \frac{\dd\ell^2(x,y)}{\ell^2(x,y)}\Big)
=\lim_{x\to y} \dd G(x,y) +\frac{{\bf 1}_{2\times 2}}{2\pi} \dd\s(y) \ ,
\ee
which shows that $\dd G(x,y)$ must have a finite limit as $x\to y$. To compute $\dd G(x,y)$ it is easiest to express $G$ as $\int SS$ and use that $\dd\St={\cal O}(m)$~:
\ba\label{Gvar}
\hskip-9.mm\dd G(x,y)&=&\dd \int\d^2 z \sqrt{g(z)} S(x,z)S(z,y) = \dd \int\d^2 z \sqrt{\hat g(z)} e^{\s(z)-\s(x)/2-\s(y)/2} \St(x,z)\St(z,y) \nonumber\\
&=&\int\d^2 z \sqrt{\hat g(z)} e^{\s(z)-\s(x)/2-\s(y)/2} \St(x,z)\St(z,y) \Big( \dd\s(z)-\frac{\dd\s(x)+\dd\s(y)}{2}\Big) +{\cal O}(m) \, .
\ea
If we let $x\to y$, integrate over $y$ and take the trace we get
\ba\label{Gvarint}
&&\hskip-2.cm\int \d^2 y \sqrt{g(y)} \, \dd \trD G(y,y) \nonumber\\
&&= \int \d^2 y \sqrt{\hat g(y)}\int \d^2 z \sqrt{\hat g(z)}e^{\s(z)+\s(y)} \trD \St(y,z) \St(z,y) \big( \dd\s(z)-\dd\s(y) \big)
 +{\cal O}(m) \nonumber\\
&&= 0  +{\cal O}(m) \ .
\ea
Note that it is only after taking the trace that the integrand is odd under exchange of $y$ and $z$. Thus
\be\label{GRintvar}
\int \d^2 y \sqrt{g(y)} \, \dd \trD G_\zeta(y)= \int \d^2 y \sqrt{g(y)} \, \dd \trD G_{\rm R}(y)=\frac{\dd A}{2\pi} \ .
\ee
Inserting this into \eqref{sgravvar8} or \eqref{sgravvar9} yields
\be\label{sgravvar10}
\dd\Big( S_{\rm grav} +\frac{1}{48\pi} S_{\rm Liouville} +\frac{m^2}{4\pi} \big( \ln\pi-1\big) A +\frac{m^2}{4} \int\sqrt{g}\, \trD G_\zeta\Big)
=  {\cal O}(m^3)\ ,
\ee
which we can immediately integrate to get 
\ba\label{sgrav11}
 S_{\rm grav}[g,\hat g]&=& -\frac{1}{48\pi} S_{\rm Liouville}[g,\hat g] +\frac{m^2}{4\pi} \big(1- \ln\pi\big) (A-A_0) \nonumber\\
 &&-\frac{m^2}{4} \int\sqrt{g}\, \trD G_\zeta[g]+ \frac{m^2}{4} \int\sqrt{\hat g}\, \trD G_\zeta[\hat g] +{\cal O}(m^3)\ .
 \ea
 We see that the order $m^2$ contribution to the effective gravitational action is entirely determined by the integral over the sphere of the Green's function at coinciding points, $G_\zeta$ or $G_{\rm R}$.
 
This result for the gravitational action looks very similar to the result one gets in the bosonic case for a massive scalar field. However, the Liouville action
has an extra factor $+\frac{1}{2}$ that  arose from a combination of a minus sign from the fermionic determinant and another factor $-\frac{1}{2}$ as explained above. Of course, this $\frac{1}{2}$ is just the central charge of the fermionic CFT. The term $\sim m^2$ is again the integral of $G_\zeta$, but now with an additional minus sign (fermions~!), as well as $\frac{1}{2}$ and the Dirac trace. But most importantly, $G_\zeta$ is the regularised Green's function of $D^2$ which is not simply the scalar Laplacian as one had for the scalar field.

To go even further and determine $G_\zeta[g]$ in terms of  $G_\zeta[\hat g]$, or equivalently $G_{\rm R}[g]$ in terms of $G_{\rm R}[\hat g]$, we could try to  ``integrate" the infinitesimal conformal variation \eqref{Gvar}. Above, in \eqref{Gvarint} we could take advantage of symmetry arguments which we are lacking at present. Note that from \eqref{SRGRrel}
we also have the relation
\be\label{GzetaGRSRrel}
m^2 \trD G_\zeta=m^2 \trD G_{\rm R} + \frac{m^2}{\pi}(\g-\ln 2) = m \trD \g_* S_{\rm R}  + \frac{m^2}{\pi}(\g-\ln 2) \ .
\ee
So instead of working with the variation of $\trD G_\zeta$ (to order $m^0$) we can look at the variation of $\trD \g_* S_{\rm R}$ (to order $m$).  But this will lead us back to the same type of expression as \eqref{Gvar}.

Instead we will use directly the equality of $G$ and $\int SS$ and use again that $\St$ is independent of the conformal factor, up to terms of order $m$ (we write $G_g(x,y)$ instead of $G[g](x,y)$, and similarly for $S$)~:
\ba\label{Gfinite}
\hskip-9.mm G_g(x,y)&=& \int\d^2 z \sqrt{g(z)} S_g(x,z)S_g(z,y) 
=  \int\d^2 z \sqrt{\hat g(z)} e^{\s(z)-\s(x)/2-\s(y)/2} \St_g(x,z)\St_g(z,y) \nonumber\\
&=&\int\d^2 z \sqrt{\hat g(z)} e^{\s(z)-\s(x)/2-\s(y)/2} \St_{\hat g}(x,z)\St_{\hat g}(z,y)+{\cal O}(m) \, , 
\ea
so that ($\St_{\hat g}=S_{\hat g}$)
\be\label{GgmoinsGghat}
G_g(x,y)-G_{\hat g}(x,y)
= \int\d^2 z \sqrt{\hat g(z)} \Big( e^{\s(z)-\s(x)/2-\s(y)/2} -1 \Big) S_{\hat g}(x,z) S_{\hat g}(z,y)+{\cal O}(m) \ .
\ee
Note that since $\dd G(x,y)$ has a finite limit as $x\to y$, the same must be true for this finite difference.
As in \eqref{GRvar} we let $x\to y$ and add the piece $\sim \s$ and take the trace to get
\ba\label{GRgmoinsGRghat}
\trD G_\zeta[g](y)-\trD G_\zeta[\hat g](y)
&=&\trD G_{\rm R}[g](y)-\trD G_{\rm R}[\hat g](y)
\nonumber\\
&&\hskip-3.cm= \int\d^2 z \sqrt{\hat g(z)} \Big( e^{\s(z)-\s(y)} -1 \Big) \trD S_{\hat g}(y,z) S_{\hat g}(z,y)
 +\frac{\s(y)}{\pi} +{\cal O}(m)  \ .
\ea
Finally, we get for the terms multiplying $-m^2/4$ in the second line of the gravitational action \eqref{sgrav11}
\ba\label{sgarv12terms}
\hskip-1.cm \int\sqrt{g}\, \trD G_\zeta[g]- \int\sqrt{\hat g}\, \trD G_\zeta[\hat g] 
 &=& \frac{1}{\pi} \int \sqrt{\hat g}\, e^{2\s}\,\s
 + \int \sqrt{\hat g}\, \big( e^{2\s}-1\big) \trD G_\zeta[\hat g] \nonumber\\
&&\hskip-3.5cm +\int d^2 y  \sqrt{\hat g(y)} \int\d^2 z\sqrt{\hat g(z)} \Big( e^{\s(y)+\s(z)}-e^{2\s(y)}\Big) \trD S_{\hat g}(y,z) S_{\hat g}(z,y) \, ,\
 \ea
 and thus
 \ba\label{sgrav13}
\hskip-1.cm S_{\rm grav}[g,\hat g]&=& -\frac{1}{48\pi} S_{\rm Liouville}[g,\hat g] +\frac{m^2}{4\pi} \big(1- \ln\pi\big) (A-A_0) 
-  \frac{m^2}{4\pi} \int \sqrt{\hat g}\, e^{2\s}\,\s\nonumber\\
 &&-\frac{m^2}{4} \int \sqrt{\hat g}\, \big( e^{2\s}-1\big) \trD G_\zeta[\hat g] \nonumber\\
&& -\frac{m^2}{4}\int d^2 y  \sqrt{\hat g(y)} \int\d^2 z\sqrt{\hat g(z)} \Big( e^{\s(y)+\s(z)}-e^{2\s(y)}\Big) \trD S_{\hat g}(y,z) S_{\hat g}(z,y) \, .
 \ea
 The last term of the first line is the characteristic term $e^{2\s}\s$ of the Mabuchi action
 \be\label{Mabuchi}
 S_{\rm M}[g,\hat g]=\frac{4}{A}\int \sqrt{\hat g} \, e^{2\s}\, \s + \ldots ,
 \ee
 Obviously, this term is local, contrary to the other terms that are also present in \eqref{sgarv12terms} and constitute the second and third lines in \eqref{sgrav13}, and that are non-local.  Despite some effort, we have not been able to reduce these terms to purely local expressions in terms of the conformal factor $\s$ or the K\"ahler potential only. Such non-local terms involving the Green's functions on the manifold also are present in the effective gravitational action for massive scalars at higher orders in $m$, starting at $m^4$. It seems that in the present case of massive Majorana fermions, such non-local terms are already present at order $m^2$. It is interesting to note, that the Mabuchi action appears at present with a coefficient $-\frac{m^2 A}{16\pi}$ while in the case of the massive scalar field one exactly obtained the opposite coefficient $+\frac{m^2 A}{16\pi}$.
 
 %%%%%%%%%%%%%%%%%%%%%%%%%%%%%%%
 \section{Discussion and outlook}
 
In these notes, we have studied to some extent the effective gravitational action for massive fermions in two dimensions.
The appropriate mass term was a Majorana type mass  term $\int \p^\dag m \g_* \p$, and the spectral analysis we performed was based on the Dirac operator $D=i\Nsl +m\g_*$ whose eigenfunctions necessarily are complex. What might have looked as a simple generalisation of the massive scalar case, actually turned out to be technically quite involved. We performed a detailed study of the corresponding Green's functions, local zeta-functions and local heat kernels of the Dirac operator $D$ and of  its square $D^2$.
In particular we studied the variations of these quantities under infinitesimal conformal rescalings of the metric, and then tried to ``integrate" these infinitesimal variations to get the finite effective gravitational action $S[g,\hat g]$. Our result was valid for non-vanishing mass in which case there is no zero-mode which otherwise we would need to exclude from the Green's functions and zeta-functions.
However, to get a proper small-mass expansion of the effective gravitational action one needs to deal with quantities that have a well-defined limit as $m\to 0$. One should thus properly take into account the zero-modes of the massless Dirac operator $i\Nsl$ and define Green's functions, zeta-functions and heat kernels with these zero-modes subtracted. Of course the number and properties of the zero-modes crucially depend on the topology of the manifold. We will come back to this question in a separate publication. Here, in our last section, we restricted ourselves to spherical topology, where $i\Nsl$ has no zero mode and we could directly perform the small-mass expansion of all our quantities. At order $m^0$, the resulting effective gravitational action displays  the well-known Liouville action, correctly with a coefficient $\frac{1}{2}$ times the one for a single scalar field, as well as a cosmological constant action proportional to the area $A$ of the manifold. At order $m^2$ we found a local contribution involving the $\int \sqrt{\hat g}\, \s e^{2\s}$ term characteristic of the Mabuchi action. This term appeared with the same coefficient as for a massive scalar field, but with the opposite sign, as one might perhaps have expected. But we also found, at this order $m^2$, several non-local terms involving the Green's functions and Green's functions at coinciding points on the manifold. Such non-local terms showed up in the scalar case only at order $m^4$, but it seems that in the fermionic case they are already unavoidable at order $m^2$.

Finally, let us note that the gravitational action for a massive Majorana fermion has also been studied in a different approach long ago \cite{Seiberg}.
There, following the DDK approach \cite{DDK}, the theory of a Majorana fermion with the mass term being gravitationally dressed by the Liouville field has been explored to some extent. It would be interesting to pursue this approach further and relate it to ours. We hope to come back to this in the future.

\vskip3.mm
%%%%%%%%%%%%%%%%%%%%%%%%%%%%%%%%%%%%%%%%%%%%%%%%%%%%%%%%%%%%
\noindent{\bf Acknowledgements}

\vskip3.mm
\noindent
H.E. is funded by the European Union's Horizon 2020 research and innovation program under the Marie Sklodowska-Curie grant agreement no 891169. 
%H.E. is also supported by the National Science Foundation under Cooperative Agreement PHY-2019786 (The NSF AI Institute for Artificial Intelligence and Fundamental Interactions, http://iaifi.org/).
 
%%%%%%%%%%%%%%%%%%%%%%%%%%%%%%%%%%%%%%%%%%%%%%%%%%%%%%%%%%%%

%%%%%%%%%%%%%%%%%%%%%%%%%%%%%%%%%%%%%%%%%%%%%%%%%%%%%%%%
\end{document}